\documentclass[sigconf]{acmart}

\AtBeginDocument{%
  }

\newcommand{\yuansong}{\textcolor[RGB]{0, 0, 0}}
\newcommand{\Final}{\textcolor[RGB]{0, 0, 0}}

\setcopyright{acmlicensed}
\copyrightyear{2018}
\acmYear{2018}
\acmDOI{XXXXXXX.XXXXXXX}
\acmConference[Conference acronym 'XX]{Make sure to enter the correct
  conference title from your rights confirmation email}{June 03--05,
  2018}{Woodstock, NY}
\acmISBN{978-1-4503-XXXX-X/2018/06}

\usepackage{multirow}
\usepackage{microtype}
\usepackage[normalem]{ulem}

\acmSubmissionID{3815}

\copyrightyear{2026}
\acmYear{2026}
\setcopyright{cc}
\setcctype{by}
\acmConference[CHI '26]{Proceedings of the 2026 CHI Conference on Human Factors in Computing Systems}{April 13--17, 2026}{Barcelona, Spain}
\acmBooktitle{Proceedings of the 2026 CHI Conference on Human Factors in Computing Systems (CHI '26), April 13--17, 2026, Barcelona, Spain}
\acmPrice{}
\acmDOI{10.1145/3772318.3790835}
\acmISBN{979-8-4007-2278-3/2026/04}

\begin{document}

\title[``Do I Trust the AI?'' Towards Trustworthy AI-Assisted Diagnosis: Understanding User Perception in LLM-Supported Reasoning]{``Do I Trust the AI?'' Towards Trustworthy AI-Assisted Diagnosis: Understanding User Perception in LLM-Supported Clinical Reasoning}


\author{Yuansong Xu}
\orcid{0009-0005-1630-6279}
\affiliation{%
\institution{School of Information Science and Technology, ShanghaiTech University}
  \city{Shanghai}
  \country{China}
}
\email{xuys2023@shanghaitech.edu.cn}

\author{Yichao Zhu}
\orcid{0009-0000-4606-1886}
\affiliation{%
\institution{School of Information Science and Technology, ShanghaiTech University}
  \city{Shanghai}
  \country{China}
}
\email{zhuych12022@shanghaitech.edu.cn}

\author{Haokai Wang}
\orcid{0009-0006-9743-907X}
\affiliation{%
\institution{School of Information Science and Technology, ShanghaiTech University}
  \city{Shanghai}
  \country{China}
}
\email{wanghk2025@shanghaitech.edu.cn}

\author{Yuchen Wu}
\orcid{0009-0005-8333-4405}
\affiliation{%
\institution{School of Information Science and Technology, ShanghaiTech University}
  \city{Shanghai}
  \country{China}
}
\email{wuych2022@shanghaitech.edu.cn}

\author{Yang Ouyang}
\orcid{0009-0000-5841-7659}
\affiliation{%
\institution{School of Information Science and Technology, ShanghaiTech University}
  \city{Shanghai}
  \country{China}
}
\email{ouyy@shanghaitech.edu.cn}

\author{Hanlu Li}
\orcid{0009-0008-9926-9857}
\affiliation{%
  \institution{Shanghai Clinical Research and Trial Center, ShanghaiTech University}
  \city{Shanghai}
  \country{China}
}
\email{lihl1@shanghaitech.edu.cn}

\author{Wenzhe Zhou}
\orcid{0009-0002-8094-7546}
\affiliation{%
  \institution{Shanghai Clinical Research and Trial Center, ShanghaiTech University}
  \city{Shanghai}
  \country{China}
}
\email{zhouwzh1@shanghaitech.edu.cn}

\author{Xinyu Liu}
\orcid{0000-0001-6566-3647}
\affiliation{%
  \institution{Shanghai Clinical Research and Trial Center, ShanghaiTech University}
  \city{Shanghai}
  \country{China}
}
\email{nightclose95@163.com}

\author{Chang Jiang}
\orcid{0000-0002-7468-3372}
\affiliation{
  \institution{Shanghai Clinical Research and Trial Center, ShanghaiTech University}
  \city{Shanghai}
  \country{China}
}
\email{cjiang_fdu@yeah.net}

\author{Quan Li}
\authornote{Corresponding Author.}
\orcid{0000-0003-2249-0728}
\affiliation{
\institution{School of Information Science and Technology, ShanghaiTech University}
  \city{Shanghai}
  \country{China}
}
\affiliation{
  \institution{Yibandao (Suzhou) Intelligent Technology Co., Ltd.}
  \city{Suzhou}
  \country{China}
}
\email{liquan@shanghaitech.edu.cn}

\renewcommand{\shortauthors}{Yuansong Xu et al.}

\begin{abstract}
Large language models (LLMs) have shown considerable potential in supporting medical diagnosis. However, their effective integration into clinical workflows is hindered by physicians' difficulties in perceiving and trusting LLM capabilities, which often results in miscalibrated trust. Existing model evaluations primarily emphasize standardized benchmarks and predefined tasks, offering limited insights into clinical reasoning practices. Moreover, research on human–AI collaboration has rarely examined physicians' perceptions of LLMs' clinical reasoning capability. In this work, we investigate how physicians perceive LLMs' capabilities in the clinical reasoning process. We designed clinical cases, collected the corresponding analyses, and obtained evaluations from physicians (N=37) to quantitatively represent their perceived LLM diagnostic capabilities. By comparing the perceived evaluations with benchmark performance, our study highlights the aspects of clinical reasoning that physicians value and underscores the limitations of benchmark-based evaluation. We further discuss the implications of opportunities for enhancing trustworthy collaboration between physicians and LLMs in LLM-supported clinical reasoning.
\end{abstract}




\maketitle

\section{Introduction} \label{Introduction}
\par With the rapid advancement of technology, large language models (LLMs) are increasingly being integrated into medical practice, demonstrating great potential in diagnostic tasks. By leveraging the capabilities of LLMs to retrieve relevant patient data and medical literature~\cite{LLM_EHR}, identify complex symptom patterns~\cite{LLM_diagonosis}, and generate personalized recommendations~\cite{LLM_personalized}, their integration into diagnostic workflows can help reduce decision-making errors, improve diagnostic efficiency, and support informed clinical decisions.

\par Despite the demonstrated effectiveness of LLMs in medical diagnosis, realizing these benefits and their full potential depends critically on effective collaboration with physicians and the integration of these tools into clinical diagnosis reasoning, \yuansong{which is defined as the iterative process of gathering evidence, forming and evaluating diagnostic hypotheses, and making treatment decisions~\cite{elstein1978medical, barrows1982clinical}.} A key factor in this collaboration is how physicians perceive and trust the results provided by LLMs~\cite{zhang2024rethinking,maity2025large}. Trust, defined as the willingness to rely on others under uncertainty~\cite{schlicker2025we}, has been shown to depend on users' perception of LLM capabilities~\cite{liu2023trustworthy,sun2024trustllm}. For example, when physicians perceive LLMs as highly capable, they tend to be more willing to adopt their suggestions, which may lead to increased trust and a higher likelihood of accepting outputs without critical evaluation~\cite{asan2020artificial,choudhury2024large}. Conversely, when LLMs are perceived as less capable, physicians may exhibit lower levels of trust and become less inclined to incorporate their suggestions into the diagnostic process~\cite{wang2021brilliant}. Therefore, a central challenge in enabling effective physician-LLM collaboration within clinical reasoning lies in the difficulty physicians face in accurately perceiving the LLMs' capabilities (e.g., due to model opacity or performance variability), which consequently hinders the development of appropriate trust in the results and ultimately limits the safe and effective integration of LLMs into clinical workflows.


\par While existing studies have demonstrated the performance of models like GPT-4~\cite{achiam2023gpt}, Claude 3~\cite{anthropic2024claude3}, and Gemini 1.5~\cite{google2024gemini}, these evaluations are often based on standardized benchmarks and predefined tasks such as question-answering or multiple-choice~\cite{jin2019pubmedqa,jin2021disease,pal2022medmcqa}. However, real-world clinical diagnosis is more complex and dynamic, involving clinical reasoning-based differential diagnosis that includes forming initial hypotheses, gathering evidence through inquiry, and updating decisions based on new information~\cite{kassirer1991learning,bowen2006educational}. \yuansong{In clinical practice, physicians evaluate not only diagnostic outcomes but also consider the reasoning process. The adoption and trust of LLMs follow the same principle—less about absolute correctness, more about clinical validity: alignment with real workflows, responsiveness to patient and clinician needs, and assurance of medical risk mitigation.} Therefore, existing metrics fail to capture the depth and dynamic nature of clinical reasoning in practice. In the absence of clinically grounded evaluation frameworks, physicians often resort to subjective judgments to assess LLM capabilities. Yet without standardized criteria and workflow, such assessments are inconsistent and unreliable. This unreliability prevents physicians from forming accurate perceptions of LLMs' diagnostic competence. Crucially, as established earlier, trust in LLMs fundamentally depends on users' perception of their capabilities~\cite{liu2023trustworthy,sun2024trustllm}. Therefore, the resulting ambiguity in assessing these capabilities hinders the calibration of appropriate trust. In high-risk clinical reasoning, where informed decision-making is critical, a lack of trust significantly limits the potential for effective physician–LLM collaboration, ultimately undermining models' capability to enhance diagnostic outcomes.

\par Although prior studies have explored various aspects of human-LLM collaboration, including information presentation, interaction\yuansong{, and explanation integration}~\cite{bansal2021does,shen2023convxai,li2025text,lee2025veriplan,wu2025cardioai,sivaraman2023ignore}, important gaps remain. First, existing research suggests that users tend to assess an LLM's capability based on the perceived value of its responses, which in turn influences their trust in the model~\cite{sun2024trustllm,de2025measuring}. LLMs that generate responses perceived as more valuable and actionable are generally considered to be more capable and trustworthy. However, in the specific context of clinical reasoning, \textbf{the criteria that define the ``value'' of an LLM response (e.g., diagnostic accuracy, clinical relevance, coherence of reasoning) remain insufficiently defined.} Second, while physicians' subjective perceptions of LLM capabilities offer valuable insights that complement benchmark-based evaluations, \textbf{the relationship between these two perspectives and potential discrepancies between them has received limited attention.} Previous research on AI-assisted decision-making suggests that effective trust calibration requires aligning user trust with specific performance indicators, such as accuracy, confidence, or consistency~\cite{goebel2018explainable,rechkemmer2022confidence,yang2023harnessing}. Such calibration enables users to determine when to rely on AI recommendations and when to exercise caution~\cite{buccinca2021trust,rastogi2022deciding,ma2023should}. However, given the unclear relationship between users' subjective assessments of model capability and objective performance metrics, it remains uncertain how such subjective perceptions can be reliably leveraged to calibrate trust in AI systems. Recognizing potential misalignments between subjective assessment and objective performance of LLM is essential, as such divergences can reveal limitations in current benchmark-based evaluations and inform the development of more robust, clinically grounded frameworks for evaluating and deploying trustworthy AI-assisted diagnostic tools.

\par In this work, we aim to understand physicians’ perceptions of LLMs’ clinical reasoning capabilities and examine their consistency with benchmark evaluations, to advance trustworthy AI-assisted diagnosis. Given that physicians often base their trust in LLMs on subjective perceptions of the model's capabilities, we aim to develop a structured evaluation approach that integrates these perceptions with objective performance metrics. This approach aims to complement traditional benchmark-based evaluations, which rely solely on objective metrics, and to better support physicians in forming accurate subjective assessments. To address these issues, we propose two research questions (\textbf{RQ1-RQ2}). \yuansong{Although physicians increasingly encounter LLM-based tools, it remains unclear whether these systems align with their clinical needs, integrate into existing workflows, or provide value from the physician's perspective. Even high-performing models may fail in practice when clinicians cannot accurately perceive their usefulness or develop appropriate trust.} To investigate how physicians perceive LLM capabilities during clinical reasoning, we propose \textbf{RQ1}: ``\textit{How do physicians perceive the value of responses and capabilities of LLMs in the clinical reasoning process?}'' To further identify potential discrepancies between subjective perceptions and objective, metric-based evaluations and to inform the development of more trustworthy LLM-supported diagnostic systems, we propose \textbf{RQ2}: ``\textit{What is the relationship between physicians' perceived capability of LLMs and the models' performance on standardized medical benchmarks?}''

\par To answer these research questions, we conducted a two-step study \Final{through an interactive system\footnote{\Final{https://github.com/yansen-xu/LLM-Supported-Clinical-Reasoning}}}. In \textbf{Step One: Case Analysis Collection}, we collaborated with physicians from our collaboration team to meticulously design nine clinical cases from different specialties. We then collected case analyses from physicians with various expertise levels and different LLMs, including the diagnostic inquiry, diagnosis conclusion, and principles of treatment. In \textbf{Step Two: Evaluation of Case Analysis}, we first invited 11 physicians to conduct a preliminary evaluation of case analysis and iteratively derive the evaluation dimensions. We then recruited 37 participants with different levels of expertise as \textit{Evaluators}, evaluating the case analyses through scoring each evaluation dimension and ranking each analysis. Through the analysis of the results, we modeled the relationship between \textit{Evaluators}' dimension scores and overall rankings, and derived the \textit{Perceived Capability Score} to quantify the perceived LLMs' capability. We also compared the \textit{Perceived Capability Score} with benchmark performance to identify differences in their evaluation focus, and found that perceived capability increased with benchmark performance at a diminishing rate rather than linearly. \yuansong{Our study provides new insights into physician–LLM collaboration within HCI and CSCW, and contributes design implications for fostering trustworthy and effective interactions. We grounded these implications in physicians' subjective perceptions of LLM capabilities, particularly factors related to trust, acceptance, and clinical appropriateness, such as how LLM outputs are adopted and integrated into diagnostic workflows.} In sum, our contribution can be summarized as follows:
\begin{itemize}
\item We conducted a two-step study in which we first constructed a set of clinical cases and collected analyses from both LLMs and human physicians to capture their clinical reasoning capabilities. In the second step, we recruited physicians to identify the evaluation dimensions underlying perceived clinical reasoning ability and to evaluate these case analyses by providing both dimension scores and overall rankings.
\item Based on the \textit{Evaluators}' dimension scores and rankings, we proposed the \textit{Perceived Capability Score}, a quantitative metric of LLM clinical reasoning from the physicians' perspective. We further compared these scores with benchmark performance to assess their consistency in representing LLM capabilities and to examine differences in the emphasized dimensions.
\item We discussed key findings regarding LLM diagnostic analyses and considered their implications for future work aimed at leveraging LLM capabilities, facilitating physician–LLM collaboration, and enhancing physicians' perception of LLMs in clinical reasoning to support trustworthy AI-assisted diagnosis.
\end{itemize}

\section{Related Work}
\subsection{Human-AI Collaboration in Clinical Decision-making}
\par With the rapid advancement and growing adoption of AI and large language models (LLMs), these technologies are being increasingly utilized in the medical domain to assist in clinical decision-making~\cite{wang2024large,goh2024large,omar2024large}. Research on human-AI collaboration in clinical settings has expanded, with a particular focus on areas such as radiology and imaging analysis~\cite{sassmannshausen2024amplifying,yildirim2024multimodal}, diagnostic reasoning support~\cite{gu2021lessons}, and treatment recommendations~\cite{wu2025cardioai}.

\par Since AI systems cannot guarantee flawless accuracy, fully automated decision-making remains problematic, particularly in high-risk medical contexts~\cite{vasey2022reporting,jin2024hidden}. As a result, research has primarily framed AI as an augmentation tool, with a focus on AI-assisted decision-making within human–AI collaboration. Recent studies have identified several modes of human–AI interaction~\cite{gomez2025human}. In studies involving AI-assisted clinical decision-making, three specific modes of collaboration have been most commonly observed: AI-first assistance mode~\cite{annarumma2019automated,xie2020chexplain,lee2024comparative,zhang2024rethinking}, AI-follow-up assistance~\cite{steiner2018impact,yildirim2024multimodal}, and Request-driven AI assistance~\cite{khosravi2024artificial,sayin2025medsyn}. In AI-first assistance mode, the AI system provides the decision-making problem and predicted outcome simultaneously, and the physician then reviews to make the final decision~\cite{steyvers2024three}. For example, Annarumma et al.~\cite{annarumma2019automated} leveraged AI to pre-analyze medical imaging data, highlighting suspicious areas for physician verification. Zou et al.~\cite{zou2024ai} proposed MedRIA, an AI-driven medical inquiry assistant that suggests potential follow-up inquiries based on collected medical information, enabling physicians to decide whether to accept or supplement them. In AI-follow-up assistance mode, the physician makes the initial judgment, and the AI subsequently offers additional suggestions to complement the physician's reasoning. For example, in pathology, AI systems are used to flag ambiguous regions that may have been overlooked during the physician's initial interpretation~\cite{steiner2018impact}. Physicians can also customize AI recommendations while navigating high-resolution tumor images to search for pathology patterns of interest~\cite{gu2023augmenting}. \yuansong{Research on clinical decision-support systems (CDSS)~\cite{wolf2025clinical} has also examined collaboration in emergency settings, emphasizing factors such as recommendation timing and team coordination efficiency. This contrasts with our work, which focuses on physicians' individual clinical reasoning processes during disease diagnosis and treatment.} In Request-driven AI assistance mode, physicians actively engage AI tools to seek reasoning support or diagnostic suggestions. It is increasingly applied to synthesize diverse clinical data and support interactive reasoning and diagnosis in complex, multi-symptom cases~\cite{khosravi2024artificial,sayin2025medsyn}.

\par Despite its potential, human–AI collaboration in clinical decision-making is often hindered by inconsistencies between users' trust in AI and its actual capabilities, leading to either over-reliance or under-reliance~\cite{panigutti2023co,kuper2025psychological} on AI, which undermines diagnostic effectiveness. Previous research has focused on how to achieve effective human–AI complementarity~\cite{bansal2021does,zhang2020effect}, particularly on strategies that guide human decision-makers in determining when to trust AI suggestions and when to approach them with caution. Trust calibration mechanisms, such as providing uncertainty estimates~\cite{ma2023should} or integrating explanations and justifications with AI outputs~\cite{metzger2024empowering,bo2025rely,okoso2025expressions}, have been employed to align users' trust with the AI system's actual performance. Building on this research, we aim to explore how physicians perceive the capability of LLM in the clinical reasoning process and the relationship between the models' performance on medical benchmarks. Our goal is to identify how AI support is perceived and relied upon during clinical reasoning, thereby enhancing the understanding and improvement of human–AI collaboration.

\subsection{Human Evaluation of LLM Outputs}
\par As artificial intelligence technologies continue to evolve, the rapid development of LLMs has raised growing interest in how to effectively evaluate their outputs across diverse tasks and scenarios. Previous research has primarily adopted metric-based approaches to evaluate LLMs, such as auditing model outputs~\cite{rastogi2023supporting}, assessing generated responses through prompt engineering~\cite{jiang2022promptmaker,mishra2025promptaid}, and comparing the performance of different models~\cite{kahng2024llm}. However, these methods often fail to align with user preferences and are insufficient for capturing the nuanced, task-specific requirements of various application domains. \yuansong{Recent work has proposed using LLMs themselves as judges to evaluate model outputs~\cite{li2025generation}. However, in high-stakes domains such as medicine, LLM-as-a-judge approaches remain problematic due to issues of hallucination, limited clinical validity, and the inability to assess feasibility or risk.} In response to these limitations, recent studies have explored interactive evaluation methods that actively involve humans in assessing LLM outputs and identifying performance differences across models~\cite{fritsch2024llmtemporalcomparator,kahng2024llm}. For example, Kim et al. proposed \textit{Evallm}~\cite{kim2024evallm}, a system that allows users to define custom evaluation criteria and iteratively refine prompts. Similarly, Shankar et al.~\cite{shankar2024validates} proposed a mixed-initiative approach that supports the co-development of evaluation criteria and validation functions, helping ensure that assessments align with human preferences.


\par In domain-specific applications, the comprehensive evaluation of LLMs often necessitates the involvement of domain experts to incorporate specialized knowledge~\cite{szymanski2024integrating,szymanski2024comparing,degachi2025towards}. Some studies have explored the adaptation of general evaluation approaches to domain-specific contexts. For example, Choi et al.~\cite{choi2024availability} examined whether ChatGPT could help patients understand kidney cancer by collecting its responses and having physicians assess them via structured questionnaires. Similarly, Cheong et al.~\cite{cheong2024artificial} explored physicians' ratings of educational materials for obstructive sleep apnoea generated using ChatGPT and Google Bard. In addition to adapting existing methods, researchers have proposed new human evaluation frameworks grounded in literature reviews and domain-specific insights. For example, Awasthi et al.~\cite{awasthi2023humanely} introduced \textit{HumanELY}, a structured framework for comprehensive human evaluation of LLM outputs, and demonstrated its application in the healthcare domain. Likewise, Szymanski et al.~\cite{szymanski2025evaluation} integrated domain expertise into the evaluation of LLMs on complex knowledge tasks to streamline expert-driven assessment workflows. In the medical domain specifically, human evaluation efforts have largely emphasized incorporating clinical expertise to assess aspects such as factual accuracy, comprehension, and reasoning. For example, Singhal et al.~\cite{singhal2023large} proposed a multi-dimensional evaluation framework tailored to clinical contexts, while Tam et al.~\cite{tam2024framework} conducted a comprehensive literature review and proposed a practical evaluation framework applicable across various medical specialties. Despite these advances, most studies primarily center on model performance, often overlooking the perspectives of physicians as end-users, particularly in terms of how they interpret, trust, and interact with model-generated outputs.

\par Inspired by prior work, our study explores the application of human evaluation of LLM-generated outputs within the context of clinical reasoning, with a particular focus on how physicians assess case analyses across multiple dimensions. This perspective offers deeper insights into how physicians perceive the capabilities of LLMs in clinical reasoning and highlights the specific criteria they consider when making evaluative judgments.

\subsection{Enhance Physicians' Perception and Trust in AI-assisted Decision Support}
\par AI is increasingly being integrated in healthcare to support clinical decision-making. However, the opaque nature of its reasoning processes often hinders physicians' understanding and can undermine their trust in AI systems~\cite{asan2020artificial}. Additionally, prior studies have pointed to a disconnect between users' perceived trust in AI and the system's actual trustworthiness. To bridge this gap, researchers have explored various strategies to improve user comprehension and promote appropriately calibrated trust in AI-assisted decision-making~\cite{molina2022ai}.

\par One line of research focuses on calibrating the confidence of LLM outputs to support more informed human decision-making and reduce the risk of errors. For example, Qin et al.~\cite{qin2024enhancing} leveraged atypical case presentations to refine the model's confidence estimates, improving their reliability. Ma et al.~\cite{ma2023should} highlighted the importance of aligning human-AI capabilities to foster appropriately calibrated trust in AI-assisted decisions. Other studies investigate how different forms of expression affect users' understanding and trust. Kim et al.~\cite{kim2024m} conducted an empirical study on the impact of uncertainty expressions in LLM outputs on user perception and trust. Similarly, Bo et al.~\cite{bo2025rely} explored various intervention strategies to improve user reliance on LLM-generated advice. Further research emphasizes the role of explanation and interaction design in shaping user trust and comprehension. For instance, Okoso et al.~\cite{okoso2025expressions} examined how different styles of AI explanations influence human decision-making. \yuansong{Sivaraman et al.~\cite{sivaraman2023ignore} examined physicians' trust of AI by having them interact with a decision support interface that provided interpretable treatment recommendations for sepsis. Their study showed that current explanations remain limited to increase clinician reliance and highlight the need for further research.} Zhang et al.~\cite{zhang2024rethinking} advocated for human-centered decision processes, arguing that AI systems should enable expert oversight rather than focus solely on optimizing end outcomes. In addition, researchers have sought to quantify and represent trust to better support human-AI collaboration. Efforts include developing comprehensive benchmarks for evaluating the trustworthiness of LLMs across multiple dimensions~\cite{sun2024trustllm}, as well as proposing metrics that integrate both model performance and human factors to facilitate effective and trustworthy collaboration~\cite{gebecscce2025quantifying}. \yuansong{For example, Qiu et al.~\cite{qiu2025quantifying} introduced benchmarks that assess LLM reasoning abilities not only through final diagnosis accuracy but also across examination recommendations, diagnostic decision-making, and treatment planning. However, such evaluations seldom consider LLMs from the physician's perspective and often overlook how clinicians perceive and adopt these models.}

\par While prior research has explored enhancing user understanding and trust in AI and LLMs through improved presentation and interaction design, quantitative approaches to evaluating trust remain limited. Existing evaluations of LLM trustworthiness are often conducted from a broad, domain-agnostic perspective, with insufficient attention to the specific requirements of clinical reasoning contexts. Building on these insights, our work aims to examine physicians' perceptions of and trust in LLM outputs within clinical reasoning scenarios. By focusing on perceived model capabilities, we seek to develop context-sensitive measures of trust that more accurately reflect clinical realities, ultimately supporting more effective human–AI collaboration.

\section{Methodology}
\par \yuansong{We conducted a two-step study.} To answer \textbf{RQ1}: physicians' perceptions of LLMs' capabilities in the clinical reasoning process, \yuansong{we constructed case analysis data and collected physicians' analysis of cases (\textbf{Step One)}. We then collected physicians' evaluations of these analyses to obtain their perceived capabilities of LLMs, analyzing} how these perceived capabilities of LLMs are represented, identifying key dimensions underlying physicians' perceptions, and estimating the relative weight of each dimension (\textbf{Step Two}). This approach enables a quantitative representation of perceived LLM capability in clinical reasoning. \yuansong{To answer \textbf{RQ2}: the relationship between the perceived capability and benchmark-based performance, we further compared the perceived capability with benchmark performance, and examined differences in dimension emphasis between them (\textbf{Step Two}).}

\subsection{Study Overview}
\par
\begin{figure*}[h]
    \centering
    \includegraphics[width=\textwidth]{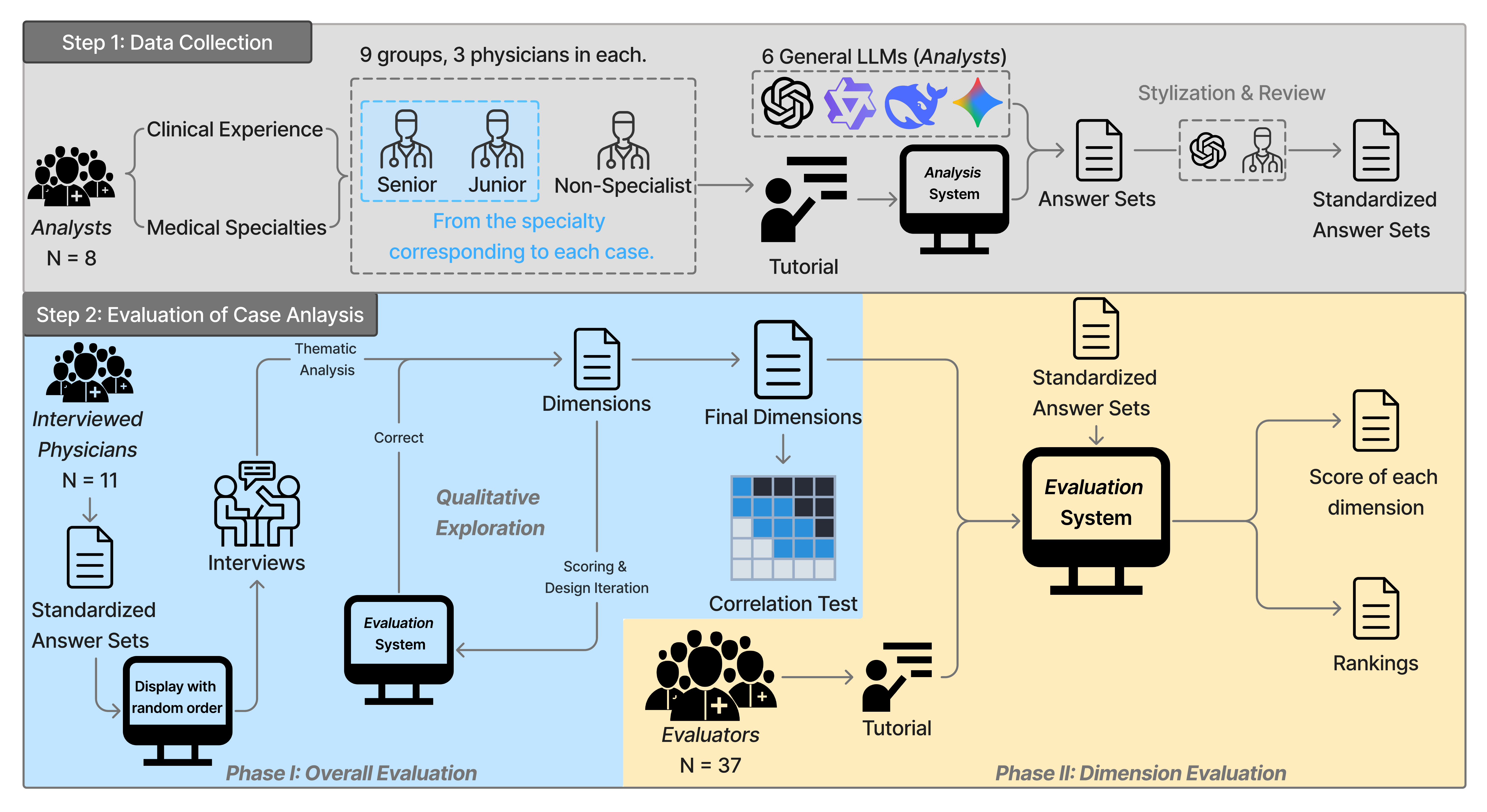}
    \caption{Step 1: Case Analysis Collection. Eight physician \textit{Analysts} were categorized by their case-specific clinical expertise into Specialist-Junior, Specialist-Senior, and Non-Specialist groups. Physicians and six general-purpose LLMs independently completed diagnostic analyses for cases. Their responses were standardized to unify textual style, and subsequently reviewed by a medical expert and the authors to form the reference answer set. Step 2: Evaluation of Case Analysis comprises two phases. In \textit{Phase I}: Overall Evaluation, we involved interviews with 11 physicians to identify evaluation dimensions. In \textit{Phase II}: Dimension Evaluation, we recruited 37 physicians as \textit{Evaluators} to evaluate and rank all \textit{Analysts}' answers across cases.}
    \label{fig:pipeline}
\end{figure*}

 \autoref{fig:pipeline} indicates the pipeline of our study: \textbf{Step One: Case Analysis Collection} and \textbf{Step Two: Evaluation of Case Analysis}. In \textbf{Step One}, we collaborated with physicians to design nine clinical cases spanning diverse specialties and difficulty levels. We recruited physicians with varying clinical backgrounds and experience, alongside multiple LLMs, to analyze the cases. Using an interactive interface, participants engaged in simulated clinical dialogues by gathering evidence, such as symptoms and test results, making inferences, and ultimately producing diagnoses and treatment plans. This process captured participants' clinical reasoning performance across the full workflow. All studies were conducted with the approval of the Institutional Review Board (IRB) and with informed consent from the participants.

\par In \textbf{Step Two}, we examined physicians' perceptions of participants' clinical reasoning capabilities using the data collected in \textbf{Step One}. This step consists of two phases. In \textit{Phase I: Overall Evaluation}, we conducted a pilot study in which physicians served as \textit{Evaluators}, providing holistic assessments of participants' performance. To further probe the rationale behind their judgments, we carried out semi-structured interviews and applied thematic analysis~\cite{braun2006using} to identify the key dimensions shaping their perceptions of capability. In \textit{Phase II: Dimensional Evaluation}, a separate group of physicians participated in the formal evaluation of the case analyses, assessing each case both holistically and across the dimensions identified in \textit{Phase I}. We then modeled the evaluation data, including both overall assessments and dimension-specific scores, to estimate the relative weight of each dimension in shaping perceived capability. This approach enabled us to quantitatively represent physicians' perceptions of LLMs' clinical reasoning capabilities.

\subsection{Step One: Case Analysis Collection}
\subsubsection{Participants and Task Materials} 
\par In Step One, both human participants and LLMs engaged in the case analysis task, and their performance data were collected. We recruited eight physicians (P1-P8) with varying clinical experience and from diverse medical specialties. Participant demographics are detailed in \autoref{tab:Case analysis participant}. Each participating physician received a \$10 compensation per case for their time and efforts. Given the specialized nature of medical diagnoses, physicians' expertise is largely confined to their own specialties; it would therefore be inappropriate to assume equal competence across other domains. Following discussions with the participating physicians, we assigned three physicians to each case, representing different expertise levels within the relevant specialty: a senior physician from the specialty (Specialist-Senior), a junior physician from the same specialty (Specialist-Junior), and a physician from another specialty (Non-specialists). This design provided a reference spectrum for benchmarking LLM performance. For instance, P5, an attending neurologist, was assigned as a senior physician for neurology cases, but due to prior rotations in internal medicine, evaluated cardiology cases as a junior physician. Detailed case assignments and physician roles are shown in \autoref{tab:assignment}. In addition, six widely used general-purpose LLMs were included in the analysis, including \textit{Deepseek-V3}, \textit{Gemini2.5pro}, \textit{GPTo3}, \textit{GPTo3-mini}, \textit{Qwen3-235b-a22b-Thinking} (abbreviated as Qwen-235B-T), and \textit{Qwq-32b} (denoted as LLM1–LLM6), each applied across all cases.


\par The diagnostic cases were constructed by five senior physicians, each with more than ten years of clinical experience. The cases cover the common initial symptom clusters in clinical practice as well as common causes encountered in medical settings. All patient information included in the case design was rigorously de-identified to ensure privacy protection. Following the problem-based learning (PBL) approach, which emphasizes clinical reasoning in medical education~\cite{xu2025advancing}, we designed nine clinical cases aimed at assessing diagnostic reasoning and analytical skills. Each case contained information such as initial symptoms, patient demographics, medical history, and ancillary examinations. \yuansong{The number of cases was determined by ensuring coverage of varying specialties and difficulty levels, while balancing the cost of case construction and the practical constraints of human evaluator workload. Besides, since our study focuses on physicians' perception of LLM capabilities, we prioritized including more evaluators to capture perceptual data, while nine cases were considered appropriate.} An example of a detailed case description is provided in the Appendix \autoref{tab:clinical case}.

\begin{table}[h]
\centering
\caption{Demographic information of human \textit{Analysts} (physicians) in \textbf{Step One: Case analysis collection.}}
\begin{tabular}{p{0.7cm} p{1.2cm} p{0.7cm} p{1.4cm} p{2.5cm}}
\hline
\textbf{ID} & \textbf{Gender} & \textbf{Age} & \textbf{Title} & \textbf{Clinical Domain} \\ \hline
P1 & Male & 33 & Attending & Orthopedics \\ \hline
P2 & Male & 26 & Intern & Orthopedics \\ \hline
P3 & Female & 31 & Resident & Breast Surgery \\ \hline
P4 & Female & 33 & Attending & Cardiology \\ \hline
P5 & Male & 30 & Resident & Neurology \\ \hline
P6 & Female & 31 & Resident & Gastroenterology \\ \hline
P7 & Female & 32 & Attending & Infectious Diseases \\ \hline
P8 & Male & 32 & Attending & Gastroenterology \\ \hline
\end{tabular}
\label{tab:Case analysis participant}
\end{table}

\begin{table}[h]
\centering
\caption{Assignment of physicians as \textit{Analysts} (P1-P9) to cases. Specialists-Senior are physicians whose specialty matches the case domain and who hold an attending (or higher) qualification; Specialists-Junior are physicians in the same specialty at the resident or equivalent junior level; Non-Specialists are physicians outside the case domain with only general medical knowledge and limited specialty expertise.}
\begin{tabular}{p{0.5cm} p{2.1cm} p{1.3cm} p{1.3cm} p{1.3cm}}
\hline
\textbf{Case} & \textbf{Specialty} & \textbf{Specialist-Senior} & \textbf{Specialist-Junior} & \textbf{Non-specialists} \\
\hline
1 & Infectious Disease & P7 & P6 & P4 \\
2 & Orthopedics & P1 & P2 & P3 \\
2 & Orthopedics & P1 & P2 & P3 \\
4 & Neurology & P5 & P7 & P4 \\
5 & Neurology & P5 & P6 & P2 \\
6 & Gastroenterology & P8 & P6 & P3 \\
7 & Cardiology & P4 & P5 & P2 \\
8 & Gastroenterology & P8 & P6 & P3 \\
9 & Cardiology & P4 & P5 & P3 \\
\hline
\end{tabular}
\label{tab:assignment}
\end{table}

\subsubsection{Procedure}
\par We designed the case analysis workflow to align with the clinical reasoning process, including \textit{evidence collection}, \textit{diagnostic reasoning}, and \textit{treatment planning}. Given that these steps are interdependent and not easily separable, a rigid stepwise structure would not ensure consistent evaluation. Instead, we implemented a dialog-based design to simulate the full clinical reasoning process, enabling users to gather evidence, request examination items and results, and formulate diagnostic and treatment decisions.

\par
\begin{figure*}[h]
    \centering
    \includegraphics[width=\textwidth]{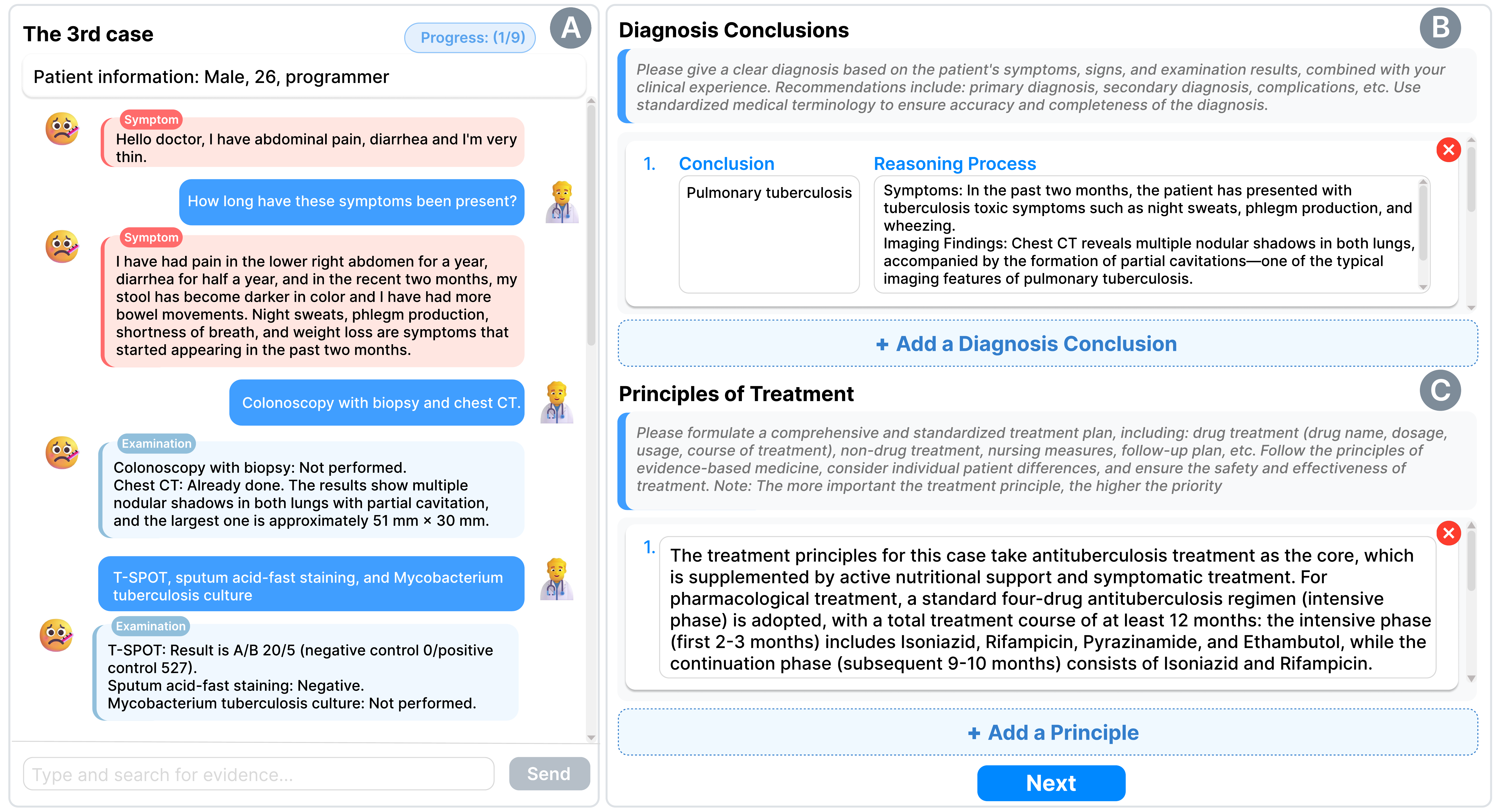}
    \caption{The interface for case analysis, including: (A) the dialogue section to obtain evidence through inquiry with the virtual patient, (B) the diagnosis section allows users to record the diagnosis conclusions with corresponding reasoning process, and (C) the treatment section allows users to record the principles of treatment.}
    \label{fig:analysis}
\end{figure*}

 For \textbf{human \textit{Analysts}}, we design an interactive interface to capture their analysis data. As shown in \autoref{fig:analysis}, the interface displays patient information in the top left corner. Participants begin with the dialogue section (\autoref{fig:analysis}-A), where they engage in inquiry with a virtual patient powered by the \textit{Qwen-plus} model~\cite{yang2025qwen3}. Carefully crafted prompts ensure that the virtual patient responds strictly based on the case background, providing examination results only when appropriate, without disclosing unsolicited information. Once \textit{Analysts} reach a diagnostic conclusion, they record it on the right side of the interface (\autoref{fig:analysis}-B), specifying the primary diagnosis, secondary diagnoses, and potential complications using standardized medical terminology. Treatment principles, rather than detailed management plans, are entered below the diagnosis (\autoref{fig:analysis}-C), covering aspects such as drug therapy, non-drug interventions, nursing care, and follow-up. This focus reflects our emphasis on the reasoning process rather than precise clinical management. After completing a case, \textit{Analysts} can save their analysis and navigate to the next case by clicking the \raisebox{-0.5ex}{\includegraphics[height=2.5ex]{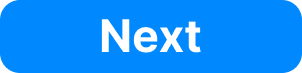}} button, which stores the dialogue content, diagnosis conclusions, and treatment principles in a JSON file. For \textbf{LLM \textit{Analysts}}, we designed prompts that guided them through inquiry, evidence collection, and formulation of diagnostic and treatment conclusions. The outputs were captured automatically through backend interactions with the models. The prompt template used for LLM case analysis is provided in the Appendix \autoref{tab:prompt}.

\subsection{Step Two: Evaluation of Case Analysis}
\par In \textbf{Step Two}, \textit{Evaluators} reviewed the case analysis data from \textbf{Step One} and evaluated their perceived capabilities across the following two phases.

\subsubsection{Phase I: Overall Evaluation}
\par In \textit{\textit{Phase I}: Overall Evaluation}, we conduct a pilot study in which physicians evaluated the collected case analysis from a holistic perspective. Using a mixed-methods design that combined qualitative exploration with quantitative validation, we identified key evaluation dimensions. We recruited 11 \textit{interviewed physicians} (E1–E11) through snowball sampling, including 6 junior physicians (1–5 years of clinical experience), 4 intermediate (6–10 years), and 1 senior physician (more than 10 years), ensuring diversity in perspectives. All \textit{Interviewed Physicians} held a medical doctorate to ensure their qualifications for evaluation. Each interviewed physician reviewed the case analyses, with both case order and analysis order randomized to mitigate order effects. During evaluation, \textit{Interviewed Physicians} engaged in think-aloud protocols and semi-structured interviews, focusing on overall impressions of capability rather than strict pairwise comparisons. They were free to examine case content while articulating the dimensions that informed their judgments. Using thematic analysis~\cite{braun2006using}, \yuansong{two authors transcribed the interview recordings. Through iterative discussion, they} identified recurring reasoning patterns and synthesized them into a preliminary code book, representing an initial set of evaluation dimensions.

\par To validate whether these candidate dimensions were both distinguishable and appropriate for evaluation, \textit{Evaluators} used the provided interface (\autoref{fig:evaluation}) to score case analyses along these dimensions and offer feedback. Based on their input, we iteratively refined the codebook \yuansong{by reviewing written comments, discussing ambiguous or overlapping dimensions, and rephrasing categories to improve clarity.} The final version, presented in \autoref{codebook}, is organized under five themes: \textit{diagnostic inquiry}, \textit{diagnosis}, \textit{clinical reasoning}, \textit{treatment}, and \textit{overall assessment}. We then conducted statistical analyses to test discriminant validity, calculating Pearson's correlation coefficients~\cite{pearson1895vii} across dimensions. As shown in \autoref{fig:dimension_pearson}, most correlations were below 0.7, indicating adequate discriminant validity. Although two pairs, e.g., diagnostic correctness vs. reasoning process (0.76), and appropriateness vs. comprehensiveness of treatment (0.83), showed higher correlations, both were retained as they capture distinct aspects: diagnostic correctness focuses on outcome, while reasoning emphasizes logic; appropriateness of treatment reflects accuracy, whereas comprehensiveness reflects coverage. Finally, internal consistency was examined using Cronbach's $\alpha$~\cite{cronbach1951coefficient} and McDonald's $\omega$. Results showed that Cronbach's $\alpha$ (0.906) and McDonald's $\omega$ (0.908) indicated excellent internal consistency across dimensions, while also suggesting partial overlap among certain dimensions, consistent with the correlation findings.


\par \textbf{Theme One: Diagnostic Inquiry.} All \textit{Interviewed Physicians} emphasized the importance of the diagnostic inquiry process, describing it as a cornerstone of modern evidence-based medicine. They highlighted that it reflects the physician's full clinical reasoning pathway: from eliciting symptoms and gathering patient information, to ordering examinations, and ultimately integrating and interpreting findings to inform the final diagnosis and treatment plan. As E1 explained, ``\textit{Just like in real diagnostic scenarios, patients do not reveal all information at once. \textit{Evaluators} must ask questions guided by their reasoning to gather further evidence for differential diagnosis.}'' Accordingly, \textit{Interviewed Physicians} stressed that a key criterion in evaluating case analyses is the logical coherence of the diagnostic inquiry process. Specifically, they assessed whether the sequence of inquiries demonstrated structured clinical reasoning, as opposed to resembling guesswork through random questioning and examinations. This underscores the importance of logical coherence in diagnostic inquiry as a distinct evaluation dimension.

\par \textbf{Theme Two: Diagnosis.} \textit{Interviewed Physicians} identified diagnosis as the core of case analysis and a key indicator of analytical capability. While some noted stylistic differences between LLMs and physicians—for example, LLMs often exhibiting greater politeness or providing more elaborate explanations—participants emphasized that the decisive factor is whether the LLM can deliver a correct diagnosis. As E5 explained, ``\textit{For patients, the most important thing is naturally whether they can receive an accurate diagnosis, as this is the prerequisite for effective treatment. If the LLM cannot provide the correct diagnosis, then other factors, such as a pleasant consultation experience, become meaningless.}'' Beyond accuracy, \textit{Interviewed Physicians} also underscored the importance of comprehensiveness in diagnostic conclusions. As E2 observed, ``\textit{In practice, patients' conditions are often complicated, so besides the main diagnosis, physicians also need to think about secondary diagnoses, possible complications, and other related factors.}'' The extent to which such considerations are incorporated was regarded as another marker of analytical competence. Drawing on these insights, we identified two evaluation dimensions within the diagnosis component: \textbf{accuracy} and \textbf{comprehensiveness} of diagnostic conclusions.

\par \textbf{Theme Three: Clinical Reasoning.} In collecting case analyses, \textit{Interviewed Physicians} were asked to document the reasoning process underlying each diagnostic conclusion. This requirement was strongly supported by the \textit{Interviewed Physicians}, who emphasized its importance for understanding how diagnostic decisions were reached. They noted that examining the soundness of the reasoning process allows for assessing \textit{Interviewed Physicians}' depth of understanding of the case and the reference points guiding their judgments. Accordingly, we identified \textbf{soundness of the reasoning process} as a key evaluation dimension.

\par \textbf{Theme Four: Treatment.} Besides diagnosis, \textit{Interviewed Physicians} emphasized that Analysts' treatment plans are also a critical indicator of capability. Although treatment accuracy often depends on diagnostic correctness, \textit{Interviewed Physicians} noted that treatment can still be evaluated independently: As E1 explained, ``\textit{Based on the patient's symptoms and condition in the current case, physicians can also propose treatment measures without considering diagnosis}''. Accordingly, we designated treatment planning as a distinct evaluation dimension, comprising two components: \textbf{appropriateness} and \textbf{comprehensiveness} of treatment principles. On one hand, treatment plans should be reasonable and align with established clinical guidelines and best practices. On the other hand, they should be comprehensive, addressing not only primary treatment but also potential complications, supportive measures, and follow-up strategies.

\par \textbf{Theme Five: Overall.} From the interviews, we learned that even if a specific diagnostic error occurs, the overall diagnostic direction and treatment principles are still considered clinically acceptable. Therefore, we adopted an overall perspective for the evaluation, which is called clinical acceptability. This perspective can be understood as the extent to which the diagnostic and treatment behaviors presented in the case analysis would be regarded as acceptable from a patient's perspective. Based on this insight, we introduced \textbf{Perceived clinical acceptability} as an overall evaluation dimension.

\par Although we identified the key dimensions for evaluating case analyses, the findings also revealed that these dimensions are not equally weighted. For example, as E5 noted, ``\textit{while the treatment plan is important, it is often determined by the diagnosis. Thus, I do not consider it a particularly critical [dimension].}'' Similarly, E3 emphasized, ``\textit{Getting the diagnosis right is the most important. The logic of the inquiry shows the reasoning process, but if the diagnosis is wrong, even the clearest inquiry logic won't help.}'' Building on these insights, we conducted \textit{Phase II: Dimensional Evaluation} to examine the relative importance of each dimension and to quantify perceived capabilities in clinical reasoning.

\begin{table*}[h]
\centering
\caption{The codebook of evaluation dimensions derived from \textit{Phase I} findings.}
\label{codebook}
\begin{tabular}{p{3cm} p{4cm} p{7cm}}
\hline
\textbf{Theme} & \textbf{Code} & \textbf{Definition} \\
\hline
\multirow{1}{*}{Diagnostic inquiry} 
 & Logical coherence of the diagnostic inquiry & Whether the inquiry process is logically structured, \yuansong{maintains clinical relevance, and includes only necessary questions that contribute to narrowing differential diagnoses.} \\
\hline

\multirow{2}{*}{Diagnosis} 
 & Accuracy of diagnostic conclusions & The extent to which the diagnostic conclusion is \yuansong{clinically correct interpretations of the case}. \\
\cline{2-3}
 & Comprehensiveness of diagnostic conclusions & Whether the diagnosis covers \yuansong{all major conditions, relevant} complications, and necessary differential diagnoses. \\
\hline

\multirow{1}{*}{Clinical reasoning} 
 & Soundness of the reasoning process & Whether the reasoning is evidence-based, \yuansong{follows a complete and traceable clinical logic}, and consistent with medical knowledge. \\
\hline

\multirow{2}{*}{Treatment} 
 & Appropriateness of treatment principles & Whether the treatment \yuansong{principles are clinically appropriate}, follows clinical guidelines and standard practice. \\
\cline{2-3}
 & Comprehensiveness of treatment principles & Whether the treatment addresses both primary and secondary problems, including complications and follow-up. \\
\hline

\multirow{1}{*}{Overall} 
 & Perceived clinical acceptability & The extent to \yuansong{to which the response is considered acceptable and actionable in real clinical practice, including the perceived coherence of its diagnostic reasoning, the feasibility of suggested actions, and the awareness of potential clinical risks.} \\
\hline
\end{tabular}
\end{table*}

\subsubsection{Phase II: Dimensional Evaluation}
\par Based on the evaluation dimensions identified in the previous phase, we invited 37 new physicians (N1-N37) with varying levels of clinical experience to assess case analyses according to these dimensions. \yuansong{The number of \textit{Evaluators} was determined with reference to typical sample sizes in HCI user studies, while ensuring the diversity in professional background and experience to reflect the heterogeneity of clinical practice.} Given physicians' heavy workloads, it was impractical to expect all \textit{Evaluators} to evaluate all cases, each of which included nine analyses. Therefore, we organized the nine cases into three groups (three cases per group) based on their specialty relevance. Each case group was targeted to be evaluated by 20 physicians, with each physician invited to evaluate at least one group. Before the experiment, they were informed of the case groups they would evaluate, but not the specific conditions represented in each case. For each case, \textit{Evaluators} provided an overall ranking while also scoring each case on the predefined dimensions. Specifically, \textit{Evaluators} were initially asked to assign quantitative scores to each analysis. However, previous studies have shown that this is cognitively demanding and strongly influenced by the presentation order~\cite{liu2024biaseye}. We then adopted a level-based approach, where \textit{Evaluators} can categorize analyses into three performance tiers (good, medium, poor), but this lacked precision, as analyses within the same tier could still differ substantially. Ultimately, \textit{Evaluators} were asked to rank the case analyses by perceived overall performance. To facilitate ranking, we retained the three-tier structure, allowing Evaluators to first categorize analyses into tiers while still emphasizing the ordering within each tier, which together produced a complete overall ranking. The collected quantitative data were used in subsequent analyses to determine the relative weight of each dimension, providing a measurable representation of how \textit{Evaluators} perceive LLM capabilities in the clinical reasoning process. Physicians received a compensation of \$25 per group evaluation. \yuansong{We also provide an example of well-performed and poorly-performed case analyses in \autoref{tab:Good_example} and \autoref{tab:Worst_example} in \autoref{AppendixF}.}

\begin{figure*}[h]
    \centering
    \includegraphics[width=\textwidth]{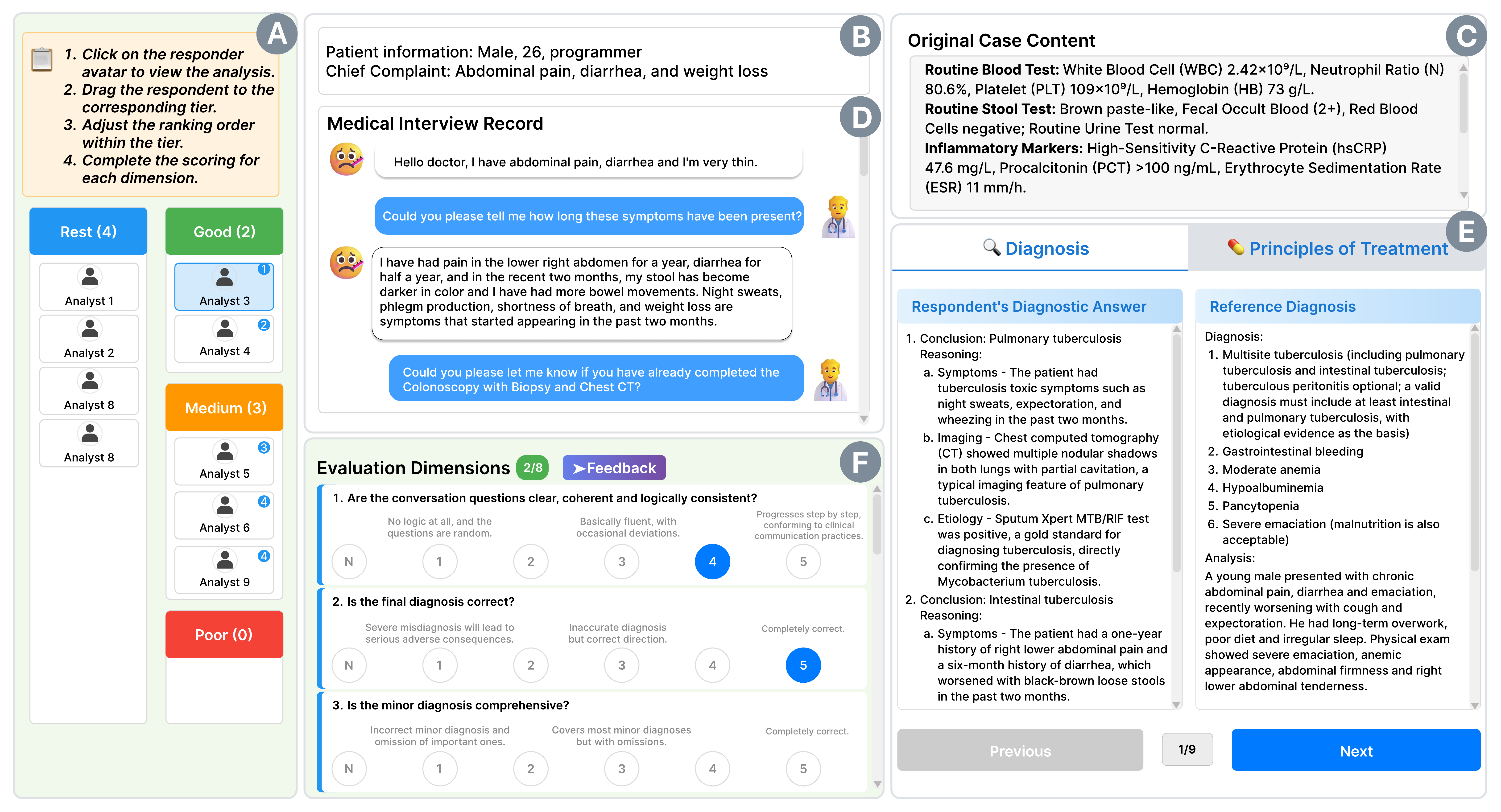}
    \caption{The interface for evaluation of case analysis, including: (A) the ranking section to assign the analyses to a specific order, (B) the basic information of current case, (C) the original case content, (D) the conversation section presents the dialogue history with the virtual patient in the current case analysis, (E) the diagnoses and treatment plans provided in current analysis alongside the reference answers, and (F) the scoring section allows users to score the analysis based on predefined evaluation dimensions.}
    \label{fig:evaluation}
\end{figure*}

\par The interface for case analysis evaluation is shown in \autoref{fig:evaluation}. Users begin with the \textbf{ranking section} (\autoref{fig:evaluation}-(A)) and follow the instructions provided in the top-left corner to complete the evaluation process. Initially, all case analyses are placed in the \raisebox{-0.5ex}{\includegraphics[height=2.5ex]{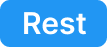}} panel. To facilitate ranking, three additional panels are provided: \raisebox{-0.5ex}{\includegraphics[height=2.5ex]{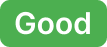}}, \raisebox{-0.5ex}{\includegraphics[height=2.5ex]{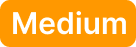}}, and \raisebox{-0.5ex}{\includegraphics[height=2.5ex]{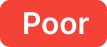}}. Users can drag analyses into the appropriate panels and then order them within each panel to generate an overall ranking.

\par In the \textbf{case section}, users can view basic information such as patient demographics and chief complaint (\autoref{fig:evaluation}-B), as well as access the original case content (\autoref{fig:evaluation}-C), which includes detailed symptoms, physical examination, and ancillary tests. Clicking on an \textit{Analysts}' avatar in the ranking section displays the corresponding case analysis. The \textbf{conversation section} (\autoref{fig:evaluation}-D) shows the diagnosis inquiry history for the selected analysis, while the \textbf{diagnosis and treatment section} (\autoref{fig:evaluation}-E) presents the diagnoses and treatment principles proposed in the analysis, alongside reference answers for consideration.

\par After reviewing the case and the analysis, users score the analysis in the \textbf{scoring section} (\autoref{fig:evaluation}-F) based on the predefined evaluation dimensions. \yuansong{We provide questions following the identified evaluation dimensions in \autoref{codebook}, including three rating anchors to facilitate \textit{Evaluators}' scoring. For example, we present the questions ``\textit{Is the final diagnosis correct}'' based on the \textit{Accuracy of diagnostic conclusions}, with three descriptions: \textit{Severe misdiagnosis and would lead to serious adverse consequences} (score = 1), \textit{Inaccurate diagnosis but correct direction} (score = 3), and \textit{Complete correct} (score = 5).} A \raisebox{-0.5ex}{\includegraphics[height=2.5ex]{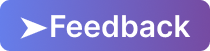}} button is also provided for users to record any additional feedback during evaluation. Once all analyses for the current case are scored, users can click the \raisebox{-0.5ex}{\includegraphics[height=2.5ex]{Img/Next.png}} button to proceed to the next case.

\section{Results}
\par To conduct a comprehensive data analysis and organize the derived results, the study is designed to answer the following analysis research questions (\textbf{ARQ1-5}). \yuansong{Theses \textbf{ARQs} are designed following the research questions \textbf{RQ1-2} in \autoref{Introduction}. For \textbf{RQ1}: physicians' perceptions of the value and capabilities of LLMs, we designed \textbf{ARQ1-3} to explore how evaluators interpret LLM-generated case analyses, how consistent their judgments are, and how different factors shape these perceptions. For \textbf{RQ2}: relationship between perceived capability and benchmark performance, we designed \textbf{ARQ4} and \textbf{ARQ5} to compare perceived capability with standardized benchmark metrics, exploring their relationships and analyzing their emphasized dimensions.}


\noindent
• \textbf{ARQ1}: How do \textit{Evaluator} characteristics and case characteristics affect the consistency of dimension scores across \textit{Evaluators}?

\noindent
- \textit{Hypothesis 1 (H1): \textit{Evaluators}' alignment with the case specialty influences their dimension scores, leading them to emphasize different dimensions for cases within versus outside their own specialty.}

\noindent
- \textit{Hypothesis 2 (H2): Case characteristics affect dimension scores, with more difficult or complex cases expected to generate greater disagreement among \textit{Evaluators}.}

\noindent
• \textbf{ARQ2}: How consistent are \textit{Evaluators}' rankings of case analyses? How do case characteristics influence this consistency?

\noindent
- \textit{Hypothesis 3 (H3): \textit{Evaluators}' rankings of case analyses will be consistent, reflecting coherent assessments of LLM outputs and alignment of physician Analysts' rankings with their predefined performance levels.}

\noindent
- \textit{Hypothesis 4 (H4): Ranking consistency is influenced by case characteristics, with more challenging cases expected to exhibit greater variability in \textit{Evaluators}' rankings.}

\noindent
• \textbf{ARQ3}: What factors influence the relationship between dimension scores and overall rankings of case analyses?

\noindent
- \textit{Hypothesis 5 (H5): The relationship of dimension scores and rankings is influenced by \textit{Evaluator}, cases, and \textit{Analysts}, such that these factors jointly determine how dimension scores translate into overall rankings.}

\noindent
• \textbf{ARQ4}: What is the relationship between LLMs' perceived capability and their benchmark performance in clinical reasoning?

\noindent
- \textit{Hypothesis 6 (H6): LLMs' benchmark performance and perceived capability are expected to be positively and linearly correlated, such that higher benchmark scores indicate greater perceived capability in clinical reasoning.}

\noindent
• \textbf{ARQ5}: Do perceived LLMs' capabilities in clinical reasoning and benchmark evaluations emphasize the same evaluation dimensions \yuansong{and how to identify them}?

\noindent
- \textit{Hypothesis (H7): Perceived capability scores and benchmark evaluations emphasize different dimensions, with benchmark metrics placing greater weight on diagnostic accuracy.}


\subsection{Overview of Data Analysis Methods}
\par \yuansong{We organized our study around five Analysis Research Questions (\textbf{ARQs}). \textbf{ARQ1–3} examine consistency in clinical reasoning dimension scores and case analysis rankings, including: (1) the impact of \textit{Evaluators} and case factors on variation and agreement in dimension-level scores (\textbf{ARQ1}); (2) the consistency of case analysis rankings across \textit{Evaluators} (\textbf{ARQ2}); and (3) the relationship between dimension scores and overall rankings, including the influence of relevant factors (\textbf{ARQ3}). \textbf{ARQ4} explores the correlation between physicians' perceived LLM capabilities and benchmark performance, while \textbf{ARQ5} compares the emphasized dimensions between perception-based and benchmark-based evaluations.}

\par \yuansong{Specifically, for \textbf{ARQ1}—dimension scoring consistency, \textit{H1} tests whether physicians' rankings of case analyses correspond to their predefined performance levels (specialists and non-specialists). Since this hypothesis compares two independent groups, we used a two-tailed independent t-test to assess significant differences in dimension scores. \textit{H2} investigates whether case complexity influences agreement among evaluators on dimension scores. Since each case was scored independently by three or more evaluators, we measured inter‑rater agreement for each dimension using Kendall's W, a non-parametric statistic based on the Friedman test ~\cite{pereira2015overview} that identifies consistent patterns in related samples~\cite{corder2009nonparametric}.}

\par \yuansong{For \textbf{ARQ2}—ranking consistency of case analyses, \textit{H3} examines whether \textit{Evaluators}' rankings of LLM-generated analyses align with predefined performance levels. Since the ranking data consists of repeated measures from the same evaluators across multiple cases, we employed the Friedman  $\chi^2$ test. \textit{H4} further investigates whether ranking consistency is affected by case difficulty or complexity. Given that consistency is an ordinal consensus measure across evaluators, we again used Kendall's W to quantify how agreement in evaluator ranking varies under different case conditions.}

\par \yuansong{For \textbf{ARQ3}—examining the relationship between dimension scores and overall rankings, \textit{H5} investigates how dimension-level scores contribute to overall rankings, considering the combined effects of evaluator, case, and \textit{Analyst} factors. To model how different evaluation dimensions contribute to overall rankings, we applied a Bradley–Terry ranking regression model since it accommodates the ordinal nature of rankings and avoids the unrealistic equal-interval assumption imposed by linear regression. Using the seven dimension scores as covariates, the model yielded weighted estimates of overall ranking quality, whose predictive validity was evaluated on the test set using Kendall’s $\tau$ coefficient, a statistic that measures the agreement between the orderings of two variables. To further account for the hierarchical and context-dependent nature of evaluators’ ranking decisions, which vary across cases, \textit{Analysts}, and their interactions, we used a cumulative link mixed model (CLMM). This method is suited for ordinal outcomes and allows us to assess how case characteristics, \textit{Analyst} performance, \textit{Evaluator} differences, and their cross-level interactions influence the mapping from dimension scores to overall rankings.}

\par \yuansong{For \textbf{ARQ4}—comparing perceived capability with benchmark performance, \textit{H6} investigates whether benchmark scores are positively correlated with physicians' perceptions of LLM capability. As both variables are continuous, we applied linear regression analysis and Pearson's r correlation to quantify the linear trend and the strength of the association.}

\par \yuansong{Finally, for \textbf{ARQ5}—examining differences in dimension emphasis between perception- and benchmark-based evaluations, \textit{H7} investigates whether user perception scores collected during interaction differ from benchmark scores in the relative weighting of reasoning dimensions, particularly whether benchmarks place greater emphasis on diagnostic accuracy. We conducted dimension-wise mean comparisons and analyzed normalized weight distribution to summarize differences in emphasized dimensions between the two evaluations, offering insights to guide iterative system design.}

\subsection{Perception of LLMs’ Capability in Clinical Reasoning}
\subsubsection{ARQ1: How do \textit{Evaluator} characteristics and case characteristics affect the consistency of dimension scores across
\textit{Evaluators}?} 
\begin{figure*}[h]
    \centering
    \includegraphics[width=\textwidth]{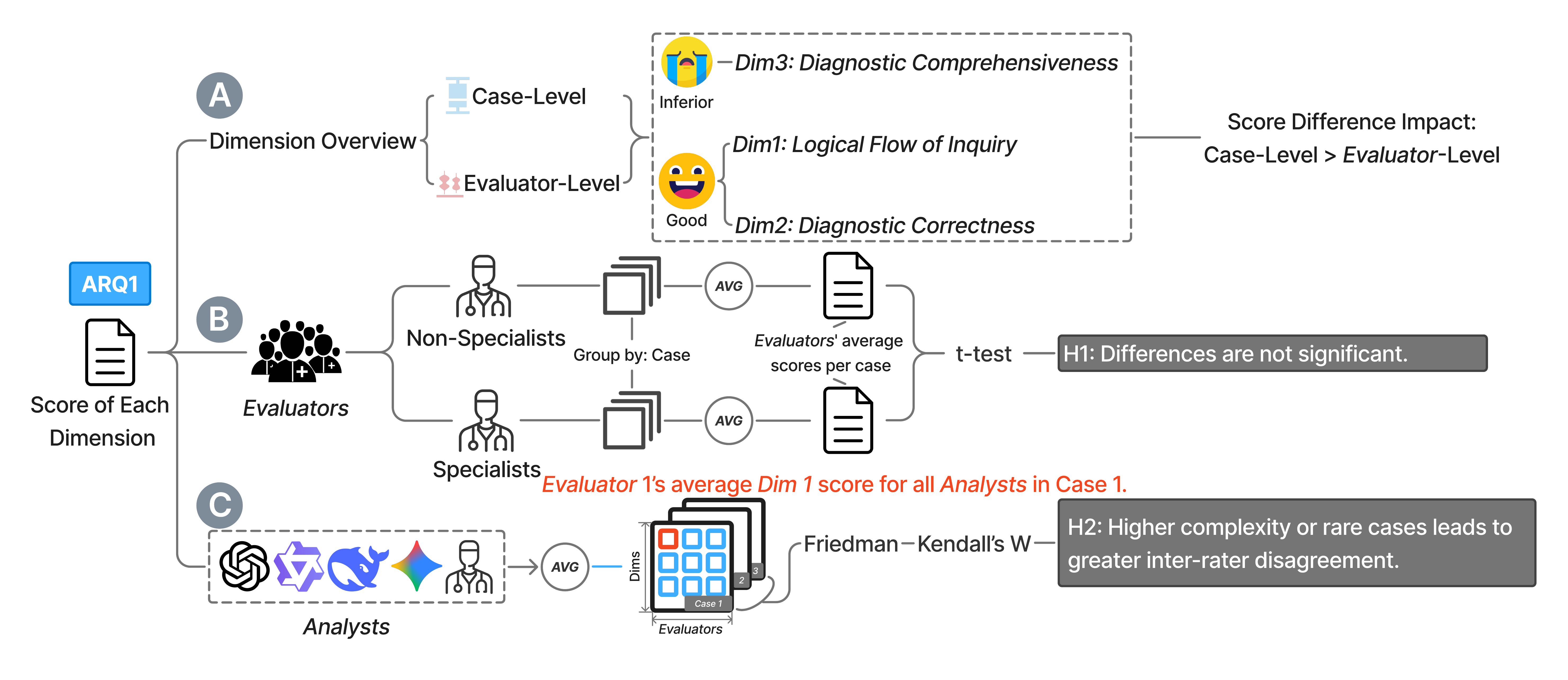}
    \caption{\yuansong{Data analysis methods for \textit{ARQ1} (dimension scores)}, including: (A) \yuansong{Overview of case-level and Evaluator-level dimension scores.} (shown in \autoref{fig:dimension_overall}). (B) Examine the differences in dimension scores between \textit{Evaluators} with two levels of expertise for each case. (C) Explore the consistency of dimension scores assigned to different \textit{Analysts} across cases.}  
    \label{fig:H1H2}
\end{figure*}

\textbf{Overall, the scores across dimensions are relatively balanced, with moderate differences. Significant variations are observed in dimension scores across physicians and cases. (H1 not supported; H2 supported)} The data analysis pipeline of the evaluation dimension scores is shown in \autoref{fig:H1H2}, including: (A) the overall distribution of dimension scores across physicians and cases, (B) the influence of \textit{Evaluators}' specialty alignment on dimension scores, and (C) the influence of case characteristics on the dimension scores.

\par
\begin{figure*}[h]
    \centering
    \includegraphics[width=\textwidth]{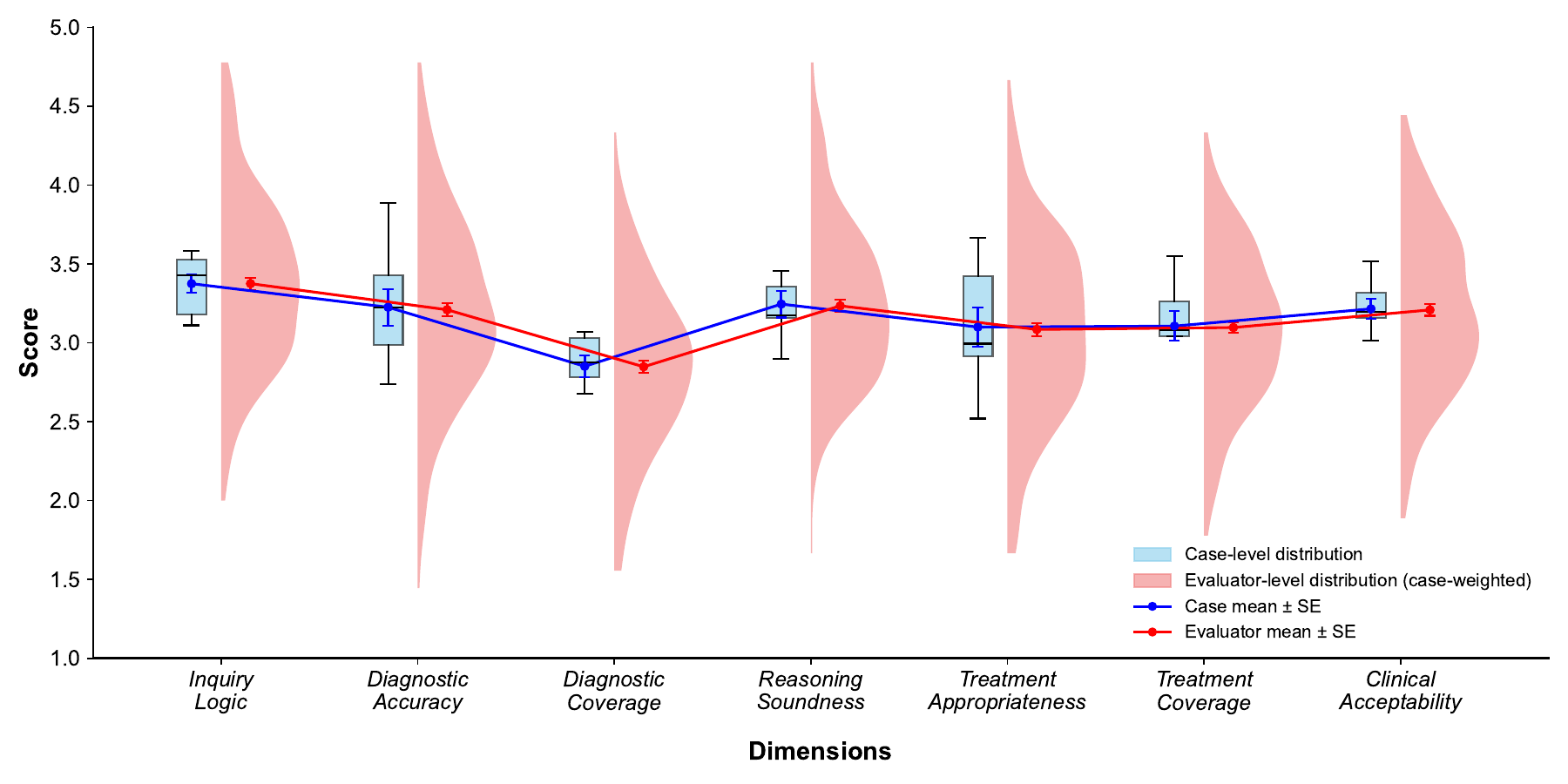}
    \caption{Distribution of dimension scores across dimensions. For each dimension, the left shows box plots of scores aggregated at the case level, and the right shows violin plots of scores aggregated at the \textit{Evaluator}-level using a case-weighted approach. Means are highlighted and connected with lines (blue: case-level, red: \textit{Evaluator}-level), with error bars showing standard errors.}
    \label{fig:dimension_overall}
\end{figure*}

 \autoref{fig:dimension_overall} presents the score distributions by dimension. The x-axis represents the evaluation dimensions, and the y-axis indicates the corresponding scores.  For each dimension, the left box plot depicts the distribution of case-level averages, while the right violin plot depicts the distribution aggregated by physician using a case-weighted approach. Because \textit{Evaluators} evaluated different numbers of cases, each \textit{Evaluator}'s scores were counted only within the cases they assessed, rather than averaged uniformly across all of their scores. Mean values are highlighted and connected with lines (blue for case-level means, red for \textit{Evaluator}-level means), with error bars denoting the standard error.

\par Across dimensions, the average scores clustered within a narrow range (2.8–3.4), indicating relatively balanced evaluations. Case-level variability is generally greater than \textit{Evaluator}-level variability, suggesting that score differences were more strongly driven by the characteristics of the case analyses than by individual \textit{Evaluators}. The dimension of \textit{Diagnostic Coverage} (Dimension 3) consistently received the lowest evaluations, whereas \textit{Inquiry Logic} (Dimension 1) and \textit{Diagnostic Accuracy} (Dimension 2) tended to score relatively higher, indicating that \textit{Analysts} struggled with diagnostic comprehensiveness but performed better in maintaining logical inquiry flow and achieving diagnostic correctness.

\par To evaluate hypothesis \textit{H1}, we categorized \textit{Evaluators} for each case into specialists (those familiar with the case's specialty) and non-specialists. We then conducted independent-samples t-tests across all dimensions and computed effect sizes (Cohen's d). As shown in \autoref{tab:t_test_cohen_d}, all t-values were negative, suggesting that specialists generally assigned slightly lower scores than the \textit{Evaluators} who are non-specialists, although these differences were not significant. \textit{Comprehensiveness of Treatment Principles} (Dimension 6) approached significance (t = -2.08, p = 0.071, d = -0.760), suggesting a relatively large practical difference. \yuansong{While} \textit{Accuracy of Diagnostic Conclusions} (Dimension 2) and \textit{Comprehensiveness of Diagnostic Conclusions} (Dimension 3) showed minimal differences (d < 0.2). Although none of the effects reached statistical significance, certain dimensions exhibited potential trends that deserve further attention.

\par To examine hypothesis \textit{H2} that dimension scores vary across cases, we assessed inter-rater consistency within each case using \textit{Kendall's W}~\cite{corder2009nonparametric}. The results (\autoref{tab:dimension_case}) reveal substantial variation in consistency across cases. In particular, cases 3, 6, and 7 all had Kendall's W values below 0.3, reflecting pronounced disagreement among \textit{Evaluators}. These cases corresponded to rare conditions in orthopedics, neurology, and gastroenterology, where higher complexity and difficulty likely contributed to the variability in scores. By contrast, other cases exhibited higher consistency. For example, case 4 (W = 0.463) demonstrated relatively aligned evaluations.

\subsubsection{ARQ2: How consistent are \textit{Evaluators}` rankings of case analyses? How do case characteristics influence this consistency?} 
\begin{figure*}[h]
    \centering
    \includegraphics[width=\textwidth]{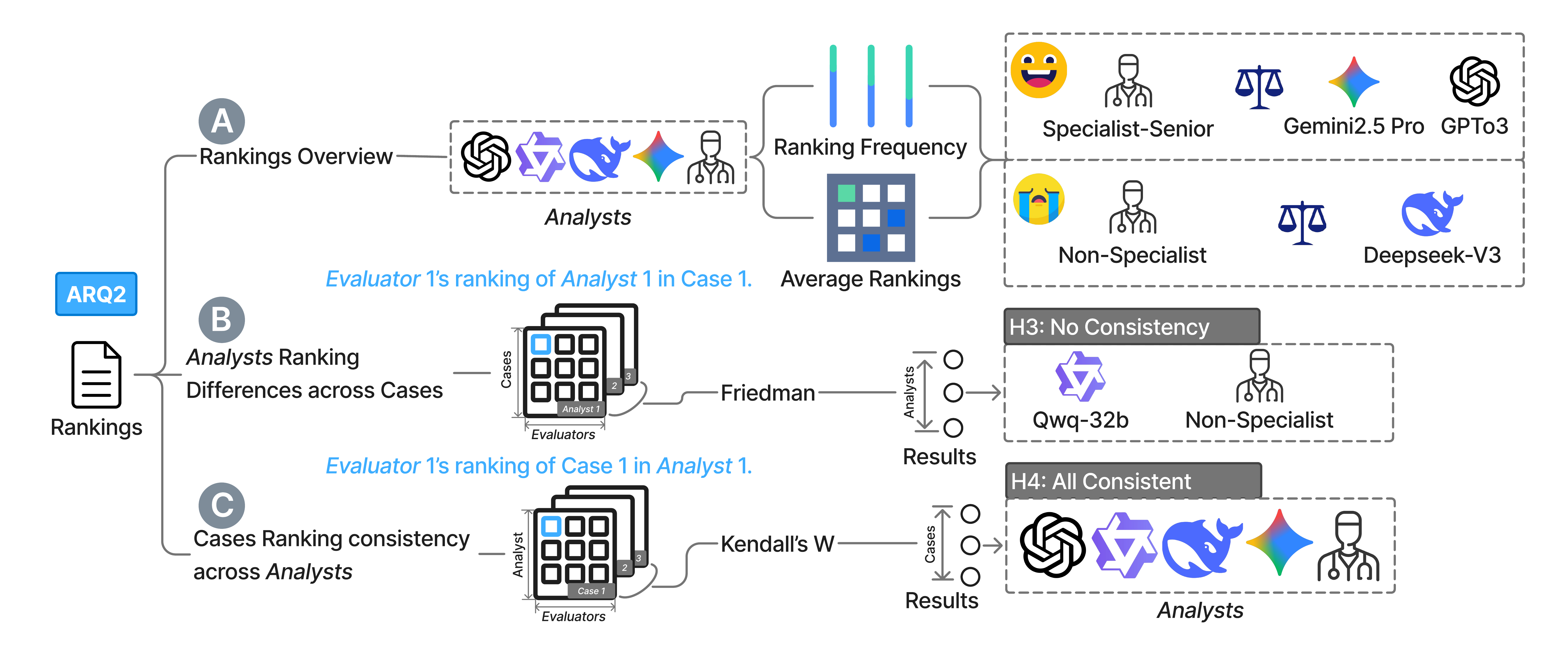}
    \caption{\yuansong{Data analysis method for ARQ2 (rankings), including: (A) \yuansong{Evaluate Analysts' performance through each analyst’s ranking frequency (\autoref{fig:Ranking_overall}-A) and rankings across cases (\autoref{fig:Ranking_overall}-B).} (B) Explore Analysts' ranking differences across cases. (C) Analyze the consistency of case rankings across Analysts}.}
    \label{fig:H3H4}
\end{figure*}

\textbf{Top-performing LLMs consistently outperformed most physicians across cases. Specialist-Senior physicians achieved comparable but less stable performance, while nearly all LLMs surpassed Non-Specialists (Hypotheses H3 and H4 partially supported).} The data analysis \yuansong{workflow} of rankings for case analysis is shown in \autoref{fig:H3H4}, including: (A) examining \textit{Analysts}' performance based on the frequency of rankings and average rankings, (B) analyzing the consistency of \textit{Evaluator}s' rankings of case analyses, and (C) analyzing the influence of case characteristics on ranking consistency. 

\begin{figure*}[h]
    \centering
    \includegraphics[width=\textwidth]{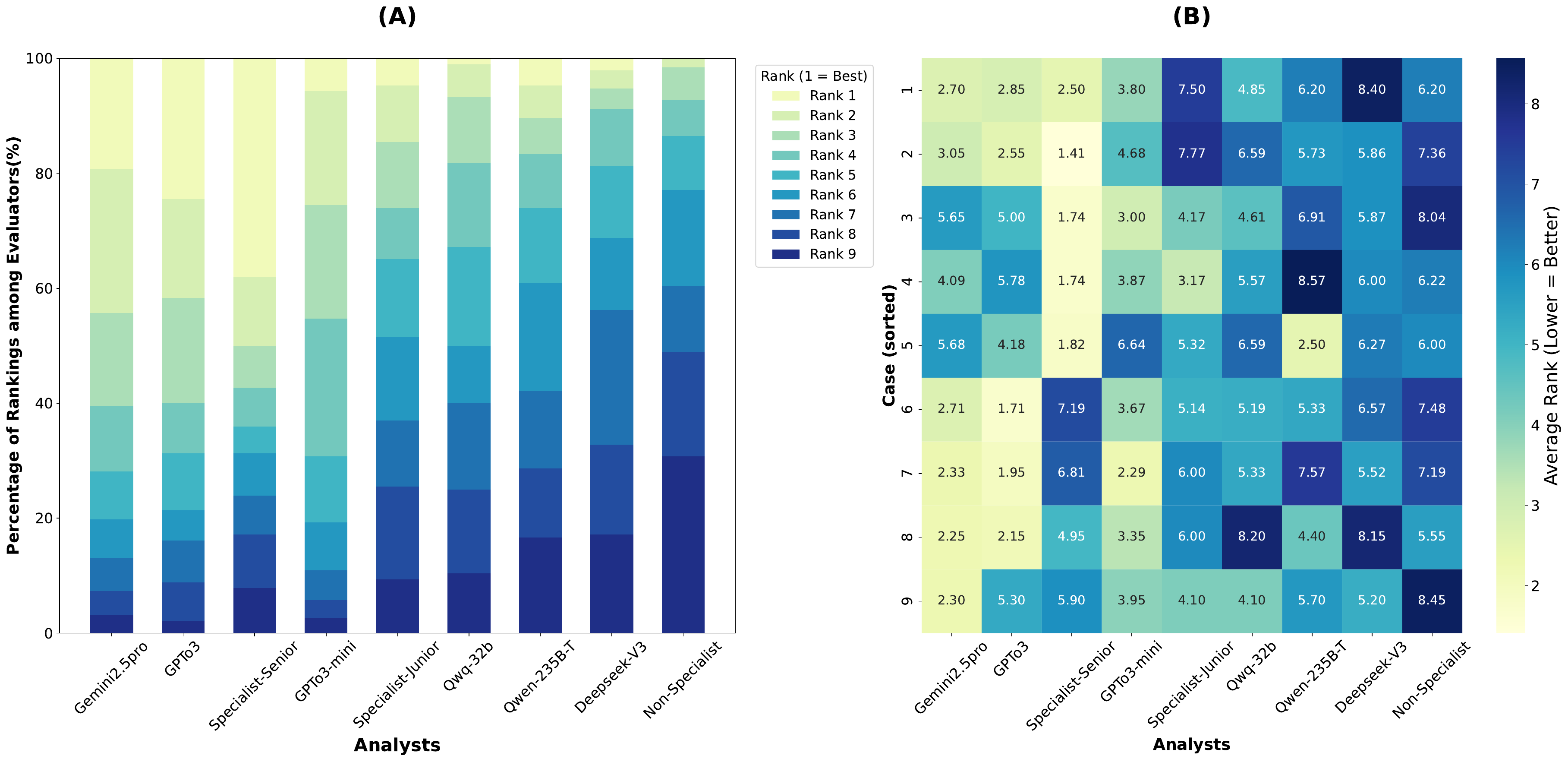}
    \caption{\yuansong{(A) The distribution of each \textit{Analyst}'s average ranking. Each vertical bar represents the ranking distribution of an Analyst across all cases. The lengths of the individual color bands correspond to the proportion of times the \textit{Analyst} received each rank, with longer and lighter bands indicating a higher proportion of top-ranked performances. \textit{Analysts} are ordered from left to right by overall performance, from best to worst. (B) The average ranking matrix of LLMs across different cases. Each column shows the average ranking of the corresponding \textit{Analyst} across cases, with lighter colors indicating higher ranks (better performance).}}
    \label{fig:Ranking_overall}
\end{figure*}

\autoref{fig:Ranking_overall}-(A) presents the rank distributions of case analyses across different LLMs and three physician groups with varying expertise levels. The y-axis indicates the percentage composition of each rank, with lighter to darker colors representing ranks 1 through 9. The results show that leading LLMs captured a higher proportion of top positions \Final{(ranks 1–3)}. \textit{Gemini 2.5 pro} and \textit{GPTo3} frequently occupied ranks 1 and 2, consistently achieving the highest evaluations. Specialist-Senior physicians also performed strongly, often appearing within the top four ranks, though their proportion of rank 1 assignments was slightly lower than that of the best-performing LLMs. In contrast, Specialist-Junior physicians exhibited mid-range distributions, similar to models such as \textit{Qwq-32b} and \textit{Qwen-235B-T}. \textit{Deepseek-V3} demonstrated comparatively weaker performance, while Non-Specialists showed the largest concentration of low ranks (7–9), reflecting the weakest overall performance.

\par We also present a heatmap of case-wise average rankings across different LLMs and physician groups in \autoref{fig:Ranking_overall}-(B). Each cell represents the average rank for a given \textit{Analysts} on a specific case, with lighter colors indicating better performance (lower rank) and darker colors indicating poorer performance (higher rank). The results reveal both substantial variation across cases and systematic differences among \textit{Analysts}. Leading LLMs such as \textit{Gemini 2.5 pro} and \textit{GPTo3} consistently occupied top positions across most cases, demonstrating strong performance and stability. Specialist-Senior physicians performed comparably to these top models but showed greater variability across cases. Specialist-Junior physicians and mid-tier models (e.g., \textit{Qwq-32b}, \textit{Qwen3-235b-a22b-Thinking}) generally held intermediate ranks. In contrast, Non-Specialists tended to rank lowest in many cases, as indicated by darker cells. Nearly all LLMs exceed the performance of Non-Specialists.

\par From \autoref{fig:Ranking_overall}, we observe that \textit{Analysts} of varying expertise levels aligned with their expected performance tiers. Specialist-Senior ranked just below the top-performing LLMs, Specialist-Junior occupied the middle range, and Non-Specialists consistently ranked below all AI models, providing partial support for hypothesis \textit{H3}. To further examine \textit{H3}, we conducted \textit{Friedman tests} to assess whether \textit{Evaluators} provided consistent rankings of the LLMs' case analyses. \yuansong{As shown in \autoref{tab:Ranking_physician}}, the results revealed significant systematic variation in rankings across cases for several groups, including \textit{Specialist-Senior} ($\chi^{2}$=22.716, p=0.004), \textit{Specialist-Junior} ($\chi^{2}$=28.604, p<0.001), and \textit{Qwen-235B-T} ($\chi^{2}$=29.138, p<0.001), as well as for leading LLMs such as \textit{GPTo3} ($\chi^{2}$=24.705, p=0.002) and \textit{Gemini 2.5 pro} ($\chi^{2}$=15.593, p=0.049). In contrast, rankings for \textit{Qwq-32b} ($\chi^{2}$=12.608, p=0.126) and \textit{Non-Specialists} ($\chi^{2}$=14.665, p=0.066) did not differ significantly across cases. These findings suggest that top-performing LLMs and more experienced \textit{Analysts} exhibited case-dependent variability in rankings, whereas less experienced physicians and certain mid-tier models produced more consistent, though generally lower, rankings across cases.


\par To evaluate hypothesis \textit{H4}, we computed \textit{Kendall's W} on the combined rankings of all \textit{Evaluator}s for each case, assessing both the consistency of rankings and the extent to which cases influenced \textit{Evaluator}s' judgments. \yuansong{As shown in \autoref{tab:case_rank}}, the results show that Kendall's W values ranged from 0.39 to 0.67 across cases. Cases 1, 2, 7, and 8 demonstrated moderate-to-high agreement ($W \approx 0.64-0.67$), cases 3, 4, and 6 reflected moderate agreement ($W \approx 0.43-0.52$), while cases 5 and 9 showed relatively weak agreement ($W \approx 0.39-0.40$). For Case 5, a rare multi-system disease, disagreement likely stemmed from differences in interpreting diagnostic clues across organ systems. For Case 9, a common cardiologic condition, variation may have reflected \textit{Evaluator}s' differing emphases on multi-dimensional clinical details. All p-values were highly significant ($p \ll 0.05$), confirming that rankings were consistently above chance. Overall, while \textit{Evaluator}s did not rank \textit{Analysts} identically across all cases, the results indicate a clear partial consensus, with the strength of agreement varying depending on case characteristics.


\subsubsection{ARQ3: What factors influence the relationship between dimension scores and overall rankings of case analyses?} 
\begin{figure*}[h]
    \centering
    \includegraphics[width=\textwidth]{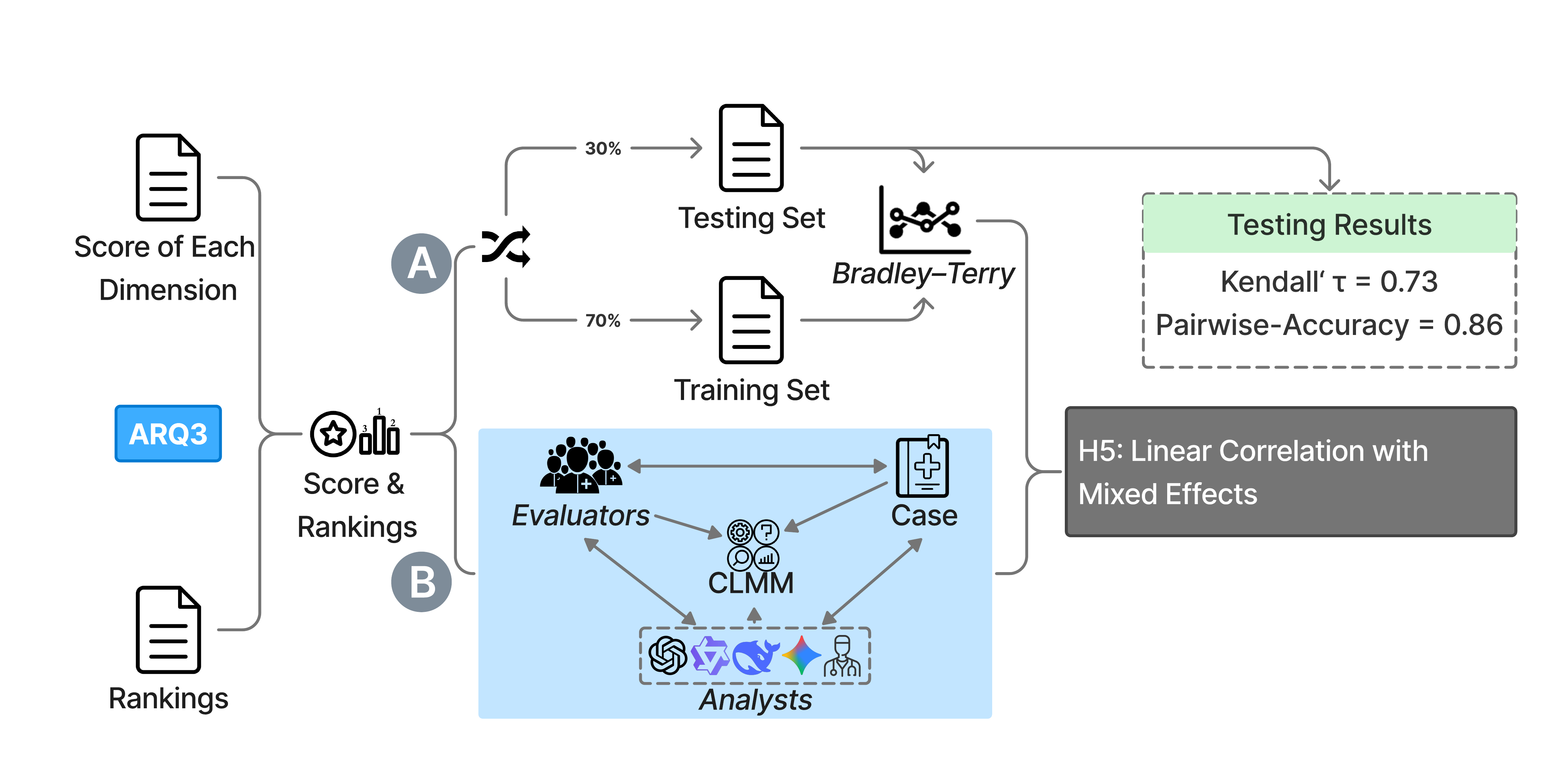}
    \caption{\yuansong{The data analysis workflow for \textit{ARQ3}, which examines the correlation between rankings and dimension scores, includes: (A) modeling the relationship between dimension scores and overall rankings, and (B) analyzing how case characteristics, \textit{Analyst} performance, Evaluator differences, and their mixed effects influence the mapping from dimension scores to overall rankings.}}
    \label{fig:H5}
\end{figure*}

\textbf{Dimension scores show a linear relationship with rankings, while the rankings themselves are further shaped by the complex, multi-level effects of \textit{Evaluators}, cases, and \textit{Analysts} (Hypothesis H5 supported).} \yuansong{We present the data analysis workflow for \textbf{ARQ3} in \autoref{fig:H5}}. To explore the relationship between the dimension scores and overall rankings, we first correlate the case analysis dimension scores with the corresponding rankings to obtain a quantitative representation. Specifically, each case included nine responses, with \textit{Evaluator}s providing both an overall ranking and independent scores across seven dimensions. A straightforward approach would be to apply linear regression, treating rankings as the dependent variable and dimension scores as predictors. However, linear regression assumes equal intervals between adjacent ranks (e.g., the difference between ranks 1 and 2 is the same as between ranks 8 and 9), which does not reflect the reality of subjective evaluations. To address this limitation, we adopted the \textit{Bradley–Terry ranking regression model}~\cite{bradley1952rank}. This model leverages pairwise comparisons to estimate the relative contribution of each dimension to the overall ranking. Incorporating the seven-dimension scores as covariates yields weighted estimates that reflect the overall quality of responses, and we define the weighted sum of these seven dimensions as the \textit{Perceived Capability Score}.

\par \yuansong{As shown in \autoref{fig:H5}-A}, we fitted a \textit{Bradley–Terry ranking regression model} using the seven-dimension scores of analyses to predict \textit{Evaluators}' rankings, with a 7:3 train–test split. The model converged with all positive weights, indicating that each dimension contributed positively to overall rankings. On the test set, the model achieved a mean Kendall's $\tau$ of 0.73, an exact match accuracy of 0.07, and a pairwise accuracy of 0.86, demonstrating strong predictive performance in capturing relative ranking patterns.

\par We had previously modeled the relationship between \textit{Evaluators}' dimension scores and rankings using the Bradley–Terry method, but effects such as cases, \textit{Analysts}, \textit{Evaluators}, and their \yuansong{mixed-effects} remained unexplored. To further investigate \textit{H5}, we employed a cumulative link mixed model (CLMM)~\cite{taylor2023rating}
 to examine how these factors influence the relationship between \textit{Evaluators}' scores and overall rankings \yuansong{(\autoref{fig:H5}-B)}. CLMMs are ideal for analyzing rank data, as they preserve category order without assuming equal intervals. Unlike traditional summaries, CLMMs account for hierarchical structures, such as evaluators assessing multiple cases and analysts, enabling more accurate modeling of dependencies and variability across levels. \Final{The model incorporates the seven dimension scores as fixed effects and treats \textit{Evaluator}, case, and \textit{Analyst} as random effects to capture hierarchical variability. In addition, pairwise interaction effects among these factors are included to account for context-dependent differences in ranking behavior.}
As reported in \autoref{tab:CLMM}, all seven dimensions significantly influenced rankings (p < 0.05). \yuansong{\textit{Perceived clinical acceptability} had the largest effect, indicating that physicians prioritize the overall actionability, feasibility, and risk awareness of responses. Diagnostic dimensions, including \textit{Inquiry Logic}, \textit{Diagnostic Accuracy}, \textit{Diagnostic Coverage}, and \textit{Reasoning Soundness}, also strongly shaped rankings, reflecting physicians' emphasize to fundamental reasoning quality. Treatment-related dimensions had smaller effects, suggesting that treatment recommendations are considered less central when forming comparative judgments.}

\par The variance decomposition (\autoref{tab:variance_decomposition}) revealed the contributions of different effects to ranking variability. The \textit{case–Analyst} interaction explained 9.3\% of variance, indicating that certain \textit{Analysts} performed particularly well or poorly on specific cases. The \textit{Evaluators–Analyst} interaction contributed 2.5\%, while the \textit{Evaluators–case} interaction was negligible, suggesting that \textit{Analyst–case} specific biases were largely captured by the fixed effects. Residual variance comprised 65.5\%, likely reflecting randomness in \textit{Evaluators}' scores due to varying expertise and unmodeled variability in \textit{Analysts'} performance.

\begin{table}[h]
\centering
\caption{Fixed effects (dimension coefficients) from the cumulative link mixed model (CLMM). (*: p <
0.05; **: p < 0.01; ***: p < 0.001)}
\label{tab:clmm_fixed}
\begin{tabular}{p{2.3cm}p{1.1cm}p{1.4cm}p{1cm}p{1.1cm}}
\toprule
\textbf{Dimension} & \textbf{Estimate} & \textbf{Std. Error} & \textbf{z value} & \textbf{p-value} \\
\midrule
Inquiry Logic  & -0.44 & 0.07 & -6.76 & $***$ \\
Diagnostic Accuracy  & -0.36 & 0.08 & -4.60 & $***$ \\
Diagnostic Coverage  & -0.23 & 0.07 & -3.50 & $***$ \\
Reasoning Soundness  & -0.33 & 0.09 & -3.84 & $***$ \\
Treatment Appropriateness  & -0.18 & 0.08 & -2.34 & $*$ \\
Treatment Coverage  & -0.26 & 0.08 & -3.41 & $***$ \\
Clinical Acceptability  & -0.58 & 0.08 & -7.02 & $***$ \\
\bottomrule
\end{tabular}
\label{tab:CLMM}
\end{table}

\begin{table*}[h]
\centering
\caption{Random effects from the cumulative link mixed model (CLMM).}
\begin{tabular}{lccccccc}
\toprule
\textbf{Random effects} & \textit{Evaluator} & case & \textit{Analyst} & \textit{Evaluator}:case & case:\textit{Analyst} & \textit{Evaluator}:\textit{Analyst} & Residual \\
\midrule
\textbf{Variance (\%)}  & 9.6\%  & 6.2\% & 6.9\%  & 0.0\%       & 9.3\%       & 2.5\%         & 65.5\% \\
\bottomrule
\end{tabular}
\label{tab:variance_decomposition}
\end{table*}

\par These results underscore the multi-level \Final{processes underlying} \textit{Evaluators}' ordinal ranking decisions. Rankings are jointly influenced by dimension scores and contextual factors, with certain dimensions, such as \textit{Perceived clinical acceptability} (Dimension 7), playing a dominant role, while others exert comparatively smaller effects. Case characteristics and analyst-specific performance further modulate the relationship between dimension scores and rankings, and although physician-specific biases exist, their impact is relatively minor. The findings highlight how rankings emerge from the \Final{combined influence of} hierarchical and contextual factors.

\subsection{Relationship between the Perceived Capability and Benchmark-based Performance}
\subsubsection{ARQ4: What is the relationship between LLMs' perceived capability and their benchmark performance in clinical reasoning?} 
\begin{figure*}[h]
    \centering
    \includegraphics[width=\textwidth]{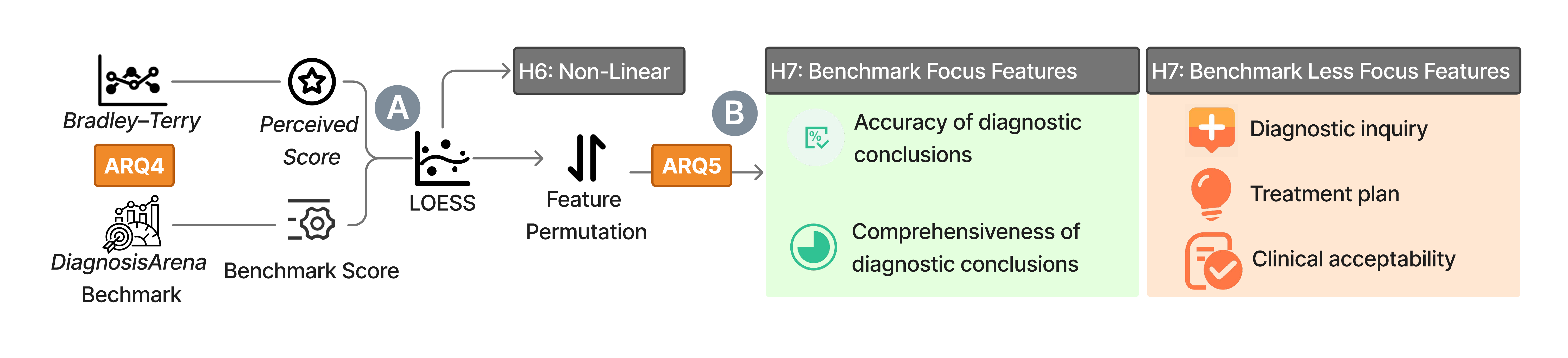}
    \caption{\yuansong{The data analysis workflow for \textbf{ARQ4} (compare \textit{Perceived Scores} with benchmark scores) and \textbf{ARQ5} (examine benchmark-relevant features), including (A) LOESS fitting to explore the relationship between LLMs' benchmark scores and their \textit{Perceived Scores} (shown in \autoref{fig:Comparation}). (B) Permutation feature-importance analysis to examine how model rankings derived from benchmark scores differ from those based on Perceived Capability Scores, identifying their difference in emphasized evaluation dimensions.}}
    \label{fig:H6H7}
\end{figure*}

\textbf{The perceived clinical reasoning capability of LLMs largely exhibited positive but not strictly linear correlation with their benchmark performance (Hypothesis H6 not supported).} \yuansong{We present the data analysis workflow of \textbf{ARQ4} in \autoref{fig:H6H7}.} To evaluate hypothesis \textit{H6}, we examined the relationship between \textit{Perceived Capability Score} and their performance on an established benchmark to identify both consistencies and divergences. We selected \textit{DiagnosisArena}~\cite{zhu2025diagnosisarena} as the benchmark. Unlike prior benchmarks based on predefined tasks such as multiple-choice questions, \textit{DiagnosisArena} evaluates diagnostic ability using real clinical cases\Final{, consistent with} our focus on clinical reasoning. \Final{In \textit{DiagnosisArena}, accuracy is computed based on whether an identical (i.e., correct) diagnosis appears among the model's top-$k$ predictions, while relevant and irrelevant diagnoses are treated as incorrect.} We adopted the reported top-1 accuracy as the benchmark metric and compared it against \textit{Perceived Capability Scores}, which are obtained by applying the weights \yuansong{obtained in ARQ3} to \textit{Evaluators}' scores \yuansong{of each dimension to obtain weighted sum,} and averaging across all case analyses for each LLM. \yuansong{We notice some models, such as \textit{DeepSeek-V3}, exhibited lower accuracy on the benchmark. On one hand, our goal is to assess a range of models and compare them with physicians of varying levels of experience; on the other hand, we found that \textit{DeepSeek-V3} still performs at the level of junior physicians in the evaluation rankings.} Given the relatively small number of models in our study (six LLMs), we employed locally weighted smoothing (LOESS)~\cite{cleveland1979robust} for non-parametric fitting \yuansong{(\autoref{fig:H6H7}-A)}. LOESS avoids the assumption of a global functional form and is better suited for capturing trends in small-sample settings, thereby enabling a more nuanced exploration of the relationship between benchmark performance and perceived capability. \yuansong{As shown in \autoref{tab:Regression}, the LOESS fit demonstrates superior performance over linear regression, exhibiting higher predictive accuracy.}

\begin{figure*}[h]
    \centering
    \includegraphics[width=\textwidth]{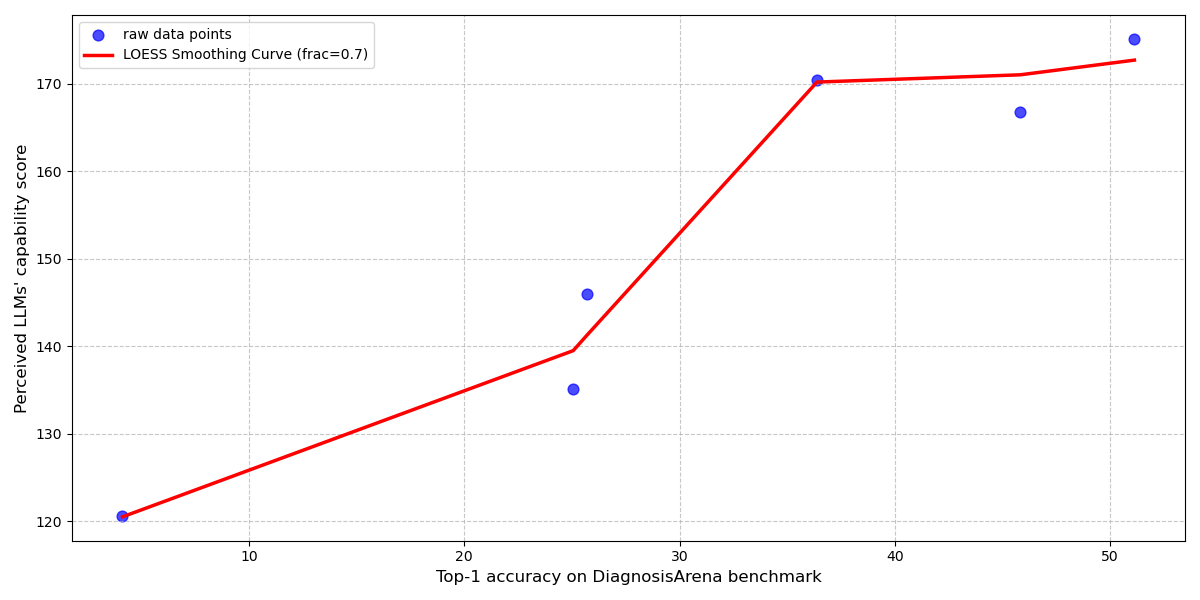}
    \caption{The results of the comparison of perceived LLMs' capabilities with benchmark-based performance.}
    \label{fig:Comparation}
\end{figure*}

\par \autoref{fig:Comparation} plots LLMs' benchmark performance (top-1 accuracy on \textit{DiagnosisArena}) against their perceived capability scores based on \textit{Evaluators}' scores. Each point represents an LLM, and the red line shows a LOESS smoothing curve. Although only six models were analyzed, the smoothed fit (R² = 0.97) indicates a strong overall correspondence between benchmark performance and perceived capability. The results reveal a generally positive trend: models with higher benchmark accuracy are typically perceived as more capable in clinical \yuansong{diagnosis}. However, the relationship does not appear strictly linear, with the slope varying and perceived gains tending to diminish at higher benchmark scores, suggesting a possible pattern of diminishing marginal returns in \textit{Evaluators}' assessments of capability.

\subsubsection{ARQ5: Do perceived LLMs' capabilities in clinical \yuansong{diagnosis} and benchmark evaluations emphasize the same evaluation dimensions \yuansong{and how to identify them}?} \textbf{Compared to \textit{Perceived Capability Scores}, benchmark evaluations place greater emphasis on the correctness and completeness of diagnostic conclusions, while giving less attention to the diagnostic inquiry, treatment principles, and perceived clinical acceptability (Hypothesis \textit{H7} supported).} To evaluate hypothesis \textit{H7}, we applied the Permutation Feature Importance method~\cite{altmann2010permutation} to examine how model rankings derived from benchmark scores differ from those based on \textit{Perceived Capability Scores}, to identify which evaluation dimensions are emphasized differently by the two approaches \yuansong{(\autoref{fig:H6H7}-B)}. For each dimension, we randomly permuted its scores across LLMs, disrupting the structure of that dimension while keeping all others unchanged. Each dimension was independently permuted 10,000 times, and the resulting permuted rankings were recorded. Benchmark rankings were derived from model performance on \textit{DiagnosisArena}, while \textit{Evaluators}' average rankings reflected the original, non-permuted perception-based rankings.

\par To quantify the impact of permutation on performance, we represented the \textit{Evaluators}’ ranking and the benchmark ranking as six-dimensional vectors, denoted as $\mathbf{R}_{\text{eval}}$ and $\mathbf{R}_{\text{bench}}$, respectively. The permuted ranking was represented as $\mathbf{R}_{\text{perm}}$. We then projected $\mathbf{R}_{\text{perm}}$ onto the line connecting $\mathbf{R}_{\text{eval}}$ and $\mathbf{R}_{\text{bench}}$, obtaining the projection point $\mathbf{P}$. The ratio of distances $\lVert \mathbf{P} - \mathbf{R}_{\text{bench}} \rVert / \lVert \mathbf{R}_{\text{eval}} - \mathbf{R}_{\text{bench}} \rVert$ was calculated, and if this ratio fell below a predefined threshold, the permuted ranking was considered substantially closer to the benchmark relative to the original \textit{Evaluators}’ ranking. Formally, let the direction vector be defined as $\mathbf{D} = \mathbf{R}_{\text{eval}} - \mathbf{R}_{\text{bench}}$, and the projection of the permuted ranking be computed as: \(\mathbf{P} = \mathbf{R}_{\text{bench}} + \frac{(\mathbf{R}_{\text{perm}} - \mathbf{R}_{\text{bench}}) \cdot \mathbf{D}}{\mathbf{D} \cdot \mathbf{D}} \mathbf{D}\),
with the projection distance defined as $d = \lVert \mathbf{P} - \mathbf{R}_{\text{bench}} \rVert$. If $d$ fell below the threshold $\epsilon$, the permutation was interpreted as shifting the ranking closer to the benchmark, indicating that the corresponding dimension may be less emphasized in benchmark evaluations. \Final{Conceptually, if permuting a dimension shifts the ranking toward the benchmark, it suggests that the benchmark underemphasizes that dimension, as such perturbations align the permuted ranking more closely with the benchmark than with perception-based evaluations.} This method highlights dimensions that are less represented in benchmark-based assessments compared to human \textit{Evaluators}' perceptions.

\begin{table*}[h]
\centering
\caption{Number of times the permuted ranking was closer to the benchmark than the \textit{Evaluators}' ranking under the defined threshold for each dimension out of 10,000 permutations.}
\begin{tabular}{c|cccc}
\hline
Dimension & Inquiry Logic & Diagnostic Accuracy & Diagnostic Coverage & Reasoning Soundness \\
\hline
Count     & 7143          & 147                 & 149                 & 121                 \\
\hline
Dimension & Treatment Appropriateness & Treatment Coverage & Clinical Acceptability & \\
\hline
Count     & 530                       & 3858              & 8065                  & \\
\hline
\end{tabular}
\label{tab:overlooked_dimension}
\end{table*}

\par As shown in \autoref{tab:overlooked_dimension}, dimensions such as \textit{Logical coherence of the diagnostic inquiry} (Dimension 1), \textit{Comprehensiveness of treatment principles} (Dimension 6), and \textit{Perceived clinical acceptability} (Dimension 7) exhibited a higher frequency of threshold-crossing changes. These findings \Final{are consistent with \textit{DiagnosisArena} benchmark's focus} on top-1 diagnostic accuracy, emphasizing diagnostic conclusions while underrepresenting other important aspects.

\subsubsection{Discussion of the comparison between the perceived capability and benchmark-based performance}
\par \yuansong{Our study compares LLM performance on a clinical diagnosis-focused benchmark with \textit{Perceived Capability Score}, aiming to reveal the emphasized different aspects between benchmark performance and perception-based judgments. We did not include other common LLM benchmarks such as multiple-choice or short-answer tests, since our focus is on models' clinical diagnosis ability. Research has also shown that models that excel on these benchmark formats do not necessarily perform well in the clinical diagnosis benchmark (\textit{DiagnosisArena})~\cite{zhu2025diagnosisarena}, underscoring their potential limitations for evaluating clinical scenarios. We acknowledge that this limits the breadth of our benchmark comparison. However, our methodology for \textbf{ARQ4} and \textbf{ARQ5} provides a quantitative comparison process to discover and understand the considerations and differences between the benchmarks and the \textit{Perceived Capability Score}. This process can also be extended to other benchmarks, providing deeper insight into how different benchmarks differ from clinicians’ perception-based assessments. These insights can help guide the design of more trustworthy AI-assisted diagnostic systems and better support physician–LLM collaboration.}

\section{Discussion}
\par We present a summary of findings on the perceived capabilities of LLM-supported clinical reasoning, highlighting both the strengths and weaknesses of LLMs compared with human physicians in practical diagnostic contexts. Building on these insights, we discuss implications for fostering trustworthy AI-assisted diagnosis and outline promising directions for future research. Finally, we acknowledge the limitations of this study.

\subsection{Summary of Key Findings}
\subsubsection{LLMs provide rapid diagnostic support but remain limited in real-world clinical integration}
\par \Final{Our study shows that LLMs can provide timely support for clinical decision-making, particularly in complex or rare cases, as their distinct reasoning patterns enable them to identify relevant conditions even with incomplete information.} Specifically, LLMs draw on statistical associations between symptoms and diseases learned from large-scale datasets, leveraging conditional probability reasoning and semantic matching~\cite{liu-etal-2024-llms}. Rather than relying on direct pathological understanding, this language- and pattern-based probabilistic reasoning enables LLMs to generate plausible diagnostic hypotheses for rare diseases under limited information.

\par Nonetheless, \Final{beyond the known constraints of traditional AI-based CDS systems,} LLMs exhibit notable limitations in practical clinical contexts. In cases requiring nuanced judgment or trade-offs, they often fail to produce actionable recommendations. As one \textit{Evaluator} (N11) observed, ``\textit{For fluid management in a patient with Acute Kidney Injury (AKI) after heart failure, the LLM suggested large-volume infusion. While this reflects partial textbook knowledge, it overlooks the full clinical context, as impaired cardiac function precludes aggressive fluid therapy, an insight derived from clinical experience.}'' In contrast, human physicians, though not always able to quantify fluid precisely, typically apply pragmatic judgment. Moreover, LLMs display limited risk assessment capabilities and occasionally propose ``textbook-style'' or unsafe treatment plans. For example, \textit{Evaluator} N3 commented, ``\textit{[LLM in the case analysis] administering high-dose diuretics to a tuberculosis patient without considering severe malnutrition is highly problematic.}'' Beyond such risks, LLMs also struggle to prioritize interventions in dynamic cases. While human physicians adjust step by step as diseases progress, LLMs tend to output broad, exhaustive treatment suggestions, many of which may be unnecessary or misaligned with the immediate clinical priority.

\subsubsection{The perceived lack of trustworthiness for LLMs often stems from the absence of diagnostic evidence acquisition processes}
\par Our study indicates that physicians' skepticism toward LLMs often arises from the absence of transparent evidence acquisition processes to support diagnostic conclusions. \Final{Whereas traditional CDS systems, while also limited in transparency, typically follow explicit rules or model outputs that constrain reasoning pathways.} While LLMs can generate correct diagnoses in many complex cases, including rare diseases, they frequently fail to provide adequate supporting rationale. This can lead to conclusions that appear less rigorous and overlook potential complications, thereby introducing clinical risks. For instance, one \textit{Evaluator} noted in a tuberculosis case: ``\textit{The LLM suggested a diagnosis of tuberculosis but did not inquire about relevant tests, leaving the recommendation clinically unverified.}'' Similarly, in a rare neurological case (Case 5), where most \textit{Analysts} were unable to reach the correct diagnosis, \textit{Gemini 2.5 Pro} successfully identified the condition. However, its reasoning process lacked coherence and could not substantiate the diagnosis with evidence\Final{, highlighting a limitation specific to LLMs' generative reasoning}. These findings suggest that in high-risk clinical scenarios, physicians are unlikely to adopt LLM-generated conclusions without transparent justification. Although LLMs can occasionally provide valuable insights that trigger ``aha moments'', their limited transparency in diagnostic reasoning significantly undermines their perceived trustworthiness as tools for AI-assisted diagnosis.

\subsubsection{Physicians' diagnostic reasoning could effectively leverage logical inference, whereas LLMs largely depend on heuristics and intuitive associations, increasing their susceptibility to diagnostic inertia}
\par Several \textit{Evaluators} observed that LLMs sometimes repeat questions during the diagnostic process or become ``stuck in a dead-end'', overly relying on a specific examination result, \Final{a pattern more pronounced in LLMs than in conventional CDS systems with rigid decision pathways.} Unlike our AI-simulated virtual patient analysis, real-world clinical practice frequently involves incomplete or uncertain patient information. In such contexts, LLMs should ideally generate reasonable inferences from the available data rather than rigidly adhering to one line of reasoning, as such rigidity can stall the diagnostic process and undermine user trust.

\par This limitation can be partly attributed to fundamental differences in diagnostic reasoning between LLMs and human physicians. Physicians' diagnostic reasoning is generally grounded in logical inference, corresponding to Kahneman's concept of slow thinking~\cite{kahneman2011fast}. When encountering conflicting evidence, they can perform systematic analysis and revise their reasoning accordingly. By contrast, LLMs generate outputs based on large-scale statistical patterns and contextual associations, more akin to heuristic or intuitive reasoning. Although techniques such as chain-of-thought prompting~\cite{wei2022chain} or structured prompts~\cite{Palagin_2023} can simulate certain aspects of slow thinking, these remain externally scaffolded rather than stemming from intrinsic cognitive control. This distinction also helps explain performance variability across cases. For example, GPT-series models showed relatively strong performance on rare cardiovascular cases, likely reflecting prior exposure to similar data rather than logical analysis or systematic reasoning.

\subsection{Broader Implication}

\par We distill the implications of our findings into three key aspects: leveraging current LLM capabilities to support physician–LLM collaboration (\ref{sec:Trust Calibration}, \ref{sec: Collaborative Deliberation}); enhancing LLMs' clinical reasoning capacity (\ref{Inquiry Workflow}, \ref{Clinical Practice and Human Experience}); and strengthening physicians' understanding and perception of LLM capabilities (\ref{LLMs' Capabilities}). \yuansong{While these implications offer design-oriented guidance from HCI and CSCW perspectives, they represent only an initial step toward comprehensive system development. Advancing physician–LLM collaboration at the system level will require substantial future work, including interface design, interaction mechanisms, and integration into clinical workflows. This study thus provides principled guidelines for the next stage of system design and implementation.}

\subsubsection{\yuansong{Support} Trust Calibration for Collaborative Diagnostic Reasoning with LLMs}\label{sec:Trust Calibration}
\par The findings of this study provide valuable insights into trust calibration in human–AI interaction. In complex, high-stakes diagnostic scenarios, collaboration cannot be reduced to a simple binary of accepting or rejecting LLM outputs. Traditional trust calibration approaches~\cite{bansal2021does,turner2022calibrating,ma2023should}, such as relying solely on a model's confidence score, are inadequate in dynamic, multifaceted contexts. Our findings indicate that trust is inherently multidimensional, with physicians dynamically weighing multiple evaluation criteria. Consequently, effective collaborative systems should expose diverse aspects of model capability rather than depend on a single metric.

\par Additionally, our study highlights discrepancies between the \textit{Perceived Capability Scores} and traditional benchmark-based evaluation. While LLMs may perform well on benchmarks, physicians' trust is influenced by factors that benchmarks often overlook, including the logical coherence of the diagnostic reasoning, consideration of clinical feasibility, and anticipation of potential risks. These findings suggest that, to foster effective physician–LLM collaboration, systems should provide transparent evidential support for diagnostic reasoning, evaluations of clinical feasibility, and predictions of intervention risks. Such features enable dynamic, context-sensitive trust calibration, ultimately promoting safer and more effective human–AI collaborative diagnostic decision-making.

\par \yuansong{\textit{Design Implication 1: Provide multi-dimensional trust calibration cues.} Systems should provide more trust calibration mechanisms to better support physicians' collaboration with LLMs. Trust calibration interventions include cues, explanations, and interactive features~\cite{wischnewski2023measuring}. Specifically, presenting uncertainty information~\cite{helldin2013presenting}, confidence levels~\cite{zhang2020effect}, and reliability updates~\cite{mcguirl2006supporting} can help users calibrate their trust appropriately. However, existing methods largely emphasize conveying the model's \textit{objective} capabilities, whereas physicians' behavior is ultimately shaped by their \textit{subjective} perception of those capabilities. This indicates the importance of explicitly incorporating perceived ability into trust calibration according to the evaluation dimensions identified in our study to enable multi-dimensional calibration. For example, the interface could present distinct, inspectable cues for the LLM's reasoning coherence and evidence completeness. For perceived clinical acceptability, we found that physicians who view an LLM's output as appropriate tend to make decisions that align with its suggestions. Systems can support trust calibration by offering feedback controls that let physicians confirm or adjust the model's reasoning path.}

\subsubsection{Aligning Physicians' and LLMs' Diagnostic Reasoning}\label{sec: Collaborative Deliberation}
\par Our study revealed that, while evaluators displayed variability in fine-grained dimension scores (ARQ1), there was considerable consensus in their overall rankings of \textit{Analysts}, indicating agreement on which analyses were superior. This high-level alignment suggests that physicians act as expert evaluators capable of critical assessment rather than passive recipients. Accordingly, collaboration between physicians and LLMs should not be conceived as a unidirectional process in which the model merely outputs results; instead, it should be structured as a ``cognitive partnership'' that encourages scrutiny, dialogue, and trust-building through interactive engagement.

\par Human–AI deliberation presents a promising framework, in which patient diagnostic reasoning can be iteratively refined through joint discussion and adjustment. Physicians can engage in \yuansong{interaction with LLMs to adjust diagnostic consideration}, focusing on critical dimensions often overlooked by \yuansong{specific} benchmark evaluations, such as the logical coherence of the diagnostic interview and the comprehensiveness of treatment planning. Both physicians and LLMs can exchange assessments, provide supporting evidence, and iteratively refine their reasoning until a final decision is reached. Inspiration can also be drawn from multidisciplinary collaboration in medicine: employing a multi-agent approach where multiple LLMs examine the same case from different perspectives, share their reasoning chains and evidential bases, and thereby support alignment between physicians' and LLMs' diagnostic reasoning to accomplish complex diagnostic tasks collaboratively.

\par \yuansong{\textit{Design Implication 2: Enable interactive, bidirectional deliberation to support reasoning alignment.} Users often encounter intent misalignment when interacting with LLMs~\cite{zhang2025neurosync}. Therefore, AI for clinical decision support should help physicians validate and calibrate its outputs~\cite{leist2025towards}, requiring systems to support collaborative reasoning where clinicians can question and refine the model’s diagnostic logic rather than passively accept it. Prior work has explored visualizations and conversational interfaces to support physician–LLM interaction and to link AI explanations to clinical evidence~\cite{samimi2025visual}. Building on this, future systems could allow physicians to highlight specific diagnosis steps and request targeted evidence or alternative reasoning paths for verification and deliberation. It is also important to leverage the complementary strengths of AI and physicians. Based on \textbf{ARQ2}, we found that these complementarities vary with the physician's background and case complexity, highlighting the need for personalized support in effective co-diagnosis. For example, AI can structure complex case reasoning and help collect and synthesize evidence to reduce omissions, while humans provide the critical reasoning, judgment, and final decision-making.}



\subsubsection{Following Physicians' Diagnostic Inquiry Workflow}\label{Inquiry Workflow}
\par In our study, \textit{Evaluators} emphasized that LLMs' diagnostic reasoning frequently lacked sufficient evidential support. This not only increased the risk of incorrect conclusions due to reasoning deviations but also diminished physicians' trust, even when the diagnostic outcomes appeared correct. These findings highlight the importance of guiding and reinforcing LLMs' diagnostic interviewing processes to ensure that conclusions and treatment suggestions are transparently grounded in evidence, and that the reasoning process coherently aligns with the final diagnosis. One feasible approach is to enhance LLMs' understanding of physicians' diagnostic workflows, from initial inquiry to final diagnosis, by collecting paired datasets of patient interviews and corresponding diagnoses and reinforcing the models with structured knowledge representations, such as knowledge graphs. This feedback-driven refinement can align LLMs' reasoning with physicians' professional logic, ultimately supporting clinical reasoning that mirrors evidence-based practices of human physicians.

\par \yuansong{\textit{Design Implication 3: Improve diagnostic reasoning through following physicians' diagnostic inquiry and clinical reasoning.} By translating clinicians' domain expertise into algorithmic strategies~\cite{liu2025scaffolded}, the system can provide more clinically grounded reasoning. When LLMs provide an initial diagnosis, the system should present the underlying evidence and structured reasoning. Consistent with clinical reasoning-based approaches to differential diagnosis~\cite{xu2025advancing}, the system can guide LLMs by prompting them to identify relevant information and provide supporting clinical guidelines or rules. Leveraging knowledge graphs and multi-agent frameworks can further support LLMs in generating diagnoses through more structured and transparent reasoning processes~\cite{zuo2025kg4diagnosis,chen2025mdteamgpt}. For example, the system could present a structured summary panel that includes the proposed diagnosis, key supporting evidence, and relevant counter-evidence.}

\subsubsection{Considering Practical Clinical Practice and Human Experience}\label{Clinical Practice and Human Experience}
\par Our findings indicate that LLMs' diagnostic and treatment recommendations must incorporate the practical rationality of clinical care, including procedural decision-making, risk assessment, and prioritization of interventions. These competencies cannot be fully acquired from textual knowledge alone, but instead require exposure to real-world clinical experience. One promising approach is to train LLMs using more sophisticated virtual patients that provide information beyond textbook scenarios—for instance, highlighting potential patient risks, prioritizing interventions based on symptom severity, and avoiding overly broad or generic responses. Additionally, \textit{Evaluators} noted that physicians' work extends beyond simply ``treating the disease''. Real clinical practice involves greater complexity, including patients' financial constraints, emotional states, and other contextual factors that influence care decisions. Current virtual patients, however, function as rational and cooperative information providers, diverging significantly from the nuances of actual patients and clinical contexts. Incorporating these factors is essential for more comprehensively evaluating and enhancing LLMs' clinical reasoning capabilities.

\par \yuansong{\textit{Design Implication 4: Integrate realistic clinical factors into system reasoning.} The system should incorporate practical clinical considerations, which can prompt clinicians to remain mindful of real-world constraints while encouraging LLMs to reason beyond idealized textbook consideration. Prior work has explored integrating AI into physician and patient knowledge structures~\cite{hao2024sdm}, supporting patient assessment and risk evaluation to enhance shared decision-making~\cite{hao2024advancing}. Future systems could provide recommendations that account not only for the diagnosis but also for contextual factors such as financial limitations, comorbid conditions, and anticipated treatment adherence, thereby aligning more closely with real-world clinical practice.}

\subsubsection{Promoting the Physicians' Perception of LLMs' Capabilities}\label{LLMs' Capabilities}
\par Prior assessments of LLM capabilities predominantly relied on benchmarks, including exam questions and case-based datasets~\cite{largelanguagemodelbenchmarks}. While exam-style benchmarks emphasize factual recall, they lack authentic clinical context and fail to measure reasoning skills. \Final{Case-based datasets are closer to real practice but still fail to capture the nuanced dynamics of physician–patient interactions.} Our study further highlights that existing benchmarks overlook LLMs' clinical acceptability, a dimension highly valued by physicians. This indicates that evaluating diagnostic correctness alone is insufficient; assessments should also consider the alignment of conclusions with patient history, examination findings, and clinical status, as well as the practical applicability of both the reasoning process and the resulting recommendations.

\par Our findings suggest that LLM evaluation should be approached not as a purely objective process but as a sociotechnical one, influenced by evaluators' expertise and the complexity of clinical cases. This aligns with the notion of situatedness~\cite{white2009sitelens}, which emphasizes that system interactions are embedded in real-world practices. Consequently, physicians should recognize the limitations of benchmark-based evaluations and adopt methods that better capture clinically relevant dimensions. For instance, a hybrid human–AI evaluation~\cite{Evalua24Human} \Final{could first have LLMs highlight pertinent dimensions,} such as clinical acceptability (e.g., diagnosing myocardial infarction as fatal or prescribing excessive diuretics for malnutrition), followed by physicians reviewing the cases based on these indicators. \yuansong{This method facilitates the explicit presentation of physicians' previously implicit overall perceptions, enabling assessments grounded in clear reasoning.}

\par \yuansong{\textit{Design Implication 5: Enable explicit, dimension-level perception capture to support system evaluation and iterative refinement.} Designers of clinical decision-support systems should incorporate interaction mechanisms that systematically capture physicians' perceptions across dimensions. By using structured tagging and maintaining traceable records—such as annotations on trust, evidence use, and feasibility—these perception data can be analyzed to inform system evaluation and iterative refinement. Leveraging methods for tracking and managing interactions~\cite{coscia2025ongoal}, the system can capture physicians’ perceptions to understand how various factors shape their views of the LLM’s capabilities, guiding improvements in physician–LLM collaborative design.}



\subsection{Limitations and Future Work}
\par This study has several limitations, which we outline across three dimensions: \textit{Experiment Settings}, \textit{Data Collection}, and \textit{Data Analysis}.

\subsubsection{Experiment Settings}
\par Within the \textit{Analysis} phase, our study design faced two key limitations. First, the pool of \textit{Analysts} was small and narrowly selected, comprising only three physicians of varying seniority per case and six general-purpose LLMs. Second, we did not include specialized medical LLMs for two main reasons: (i) many state-of-the-art systems, such as \textit{Med-PaLM}, \textit{Med-PaLM-2}~\cite{Med-PaLM-2}, and \textit{Dragon Copilot}~\cite{microsoft_dragon_copilot}, are not fully open-sourced, restricting direct API access for evaluation; and (ii) although several open-source foundation models fine-tuned on medical datasets exist (e.g., \textit{MEDITRON-7B}~\cite{Meditron_article}, \textit{HuatuoGPT}~\cite{huatuogpt}), our limited computational resources precluded local deployment and systematic testing. \yuansong{Besides, our study aims to investigate how physicians perceive LLMs' capabilities rather than comparing with state-of-the-art medical models. General-purpose LLMs such as \textit{Gemini 2.5 pro} have also shown performance comparable to senior clinicians, making them suitable for our study. Future work could involve a more comprehensive evaluation of models to gain deeper insights into perceived capabilities.} Moreover, the clinical cases we constructed remain insufficient to capture the full complexity and variability of real-world practice. Future work should address these gaps by involving a larger and more diverse set of clinical experts, incorporating specialized medical LLMs, and expanding evaluations to cover more complex and varied clinical scenarios.

\subsubsection{Data Collection}
\par In \textbf{Step One: Case Analysis Collection}, the virtual patient served primarily as a highly rational and cooperative report provider rather than a realistic patient exhibiting unexpected or nuanced behaviors. Consequently, the collected data may not fully capture the complexities of real-world clinical \yuansong{interactions}, underscoring the need for more advanced prompt engineering and intelligent virtual patient design. Future evaluations of LLMs should therefore be conducted in more naturalistic and \yuansong{clinically realistic} settings that better reflect the diverse demands physicians encounter when assessing LLM capabilities across clinical scenarios. \yuansong{Additionally, we only incorporate nine cases with eight human \textit{Analysts}, which could limit the robustness of the human performance. Future work could involve a larger number of cases and \textit{Analysts} for more comprehensive data collection and analysis.} In \textbf{Step Two: Evaluation of Case Analysis}, although we established scoring anchors for each dimension and clarified their definitions to all \textit{Evaluators}, the results may still have been subject to scale effects. Variability in physicians' subjective scoring tendencies, coupled with differences in their clinical experience and diagnostic reasoning skills, could have introduced bias into the evaluations. To address these challenges, future work could employ longitudinal study designs that allow for repeated and extended assessments, thereby providing a more comprehensive evaluation of perceived capabilities.

\subsubsection{Data Analysis}
\par Our predefined dimensions only partially captured physicians' perceptions of LLMs and were limited in addressing the complexity of real-world clinical contexts—for example, nuanced communication, disciplinary differences, or patient preferences. \yuansong{The results of \textbf{ARQ3} also indicate this: a significant portion of the variance in the rankings remained unexplained (residual). These may include individual cognitive styles and experiences of \textit{Evaluators} or contextual factors that were not captured in our study design.} Future work should move beyond fixed, generic dimensions toward developing specialized and customizable evaluation frameworks that reflect the dimensions physicians themselves prioritize in practice, thereby offering a more accurate representation of how LLM capabilities are perceived.

\section{Conclusion}
\par In this work, we seek to advance the trustworthiness of AI-assisted diagnosis in clinical reasoning by reframing the evaluation of LLM-assisted diagnosis through the lens of physicians’ perceived capabilities. \Final{Our findings reveal that clinicians attend to multiple nuanced dimensions of reasoning quality that are not fully captured by conventional metrics. By quantitatively linking dimensional judgments to holistic evaluations, we provide an interpretable framework for understanding how physicians form trust in AI-generated analyses. These insights highlight the limitations of relying solely on standardized benchmarks and underscore the importance of perception-aware evaluation in clinical AI. Our approach offers a pathway toward designing diagnostic systems and assessment protocols that better align technical performance with clinical expectations, ultimately supporting more effective and trustworthy physician–LLM collaboration in real-world practice.}

\begin{acks}
We gratefully acknowledge the anonymous reviewers for their insightful feedback. This research was supported by the National Natural Science Foundation of China (No. 62372298), the Shanghai Engineering Research Center of Intelligent Vision and Imaging, the Shanghai Frontiers Science Center of Human-centered Artificial Intelligence (ShangHAI), and the MoE Key Laboratory of Intelligent Perception and Human-Machine Collaboration (KLIP-HuMaCo).
\end{acks}

\balance
\bibliographystyle{ACM-Reference-Format}
\bibliography{sample-base}

\clearpage
\onecolumn
\appendix
\section{Example of clinical case used for Step One}
\begin{table}[h]
\centering
\caption{An example of a clinical case used for Step One: Case analysis collection in Study One.}
\label{tab:clinical case}
\begin{tabular}{p{4cm} p{10cm}}
\hline
\textbf{Initial Symptoms} & Hip pain on both sides, difficulty walking. \\
\hline
\textbf{Patient Demographics} & Female, 56 years old. \\
\hline
\textbf{Medical History} & 
\begin{itemize}
    \item 5 years ago: Onset of bilateral hip pain, more pronounced on the left side; diagnosed with bilateral femoral head necrosis via X-ray; history of steroid use for fever and fatigue 8 years ago.
    \item Pain worsened after one year of conservative treatment with traditional Chinese medicine, resulting in limited mobility.
    \item Development of subcutaneous masses with ulceration around the right hip joint, discharging milky fluid containing white granular material; wounds healed after dressing changes.
    \item Similar ulcerative lesions appeared on the right shoulder, left knee, and lumbosacral region.
    \item Currently taking oral analgesics intermittently; mobility remains limited.
\end{itemize} \\
\hline
\textbf{Auxiliary Examinations} & 
\textbf{Preoperative Imaging:}
\begin{itemize}
    \item X-ray: Lobulated masses in the periarticular soft tissue.
    \item MRI: T1WI shows markedly heterogeneous low signal; T2WI shows heterogeneous high signal; capsule exhibits long T1 and long T2 signals; surrounding soft tissue edema; contrast enhancement reveals ring-shaped enhancement of the lesion.
\end{itemize}
\textbf{Laboratory Tests:}
\begin{itemize}
    \item ESR: 76 mm/h (↑)
    \item CRP: 10.0 (↑)
    \item RF: 17 (↑)
    \item Serum calcium: Normal
    \item Serum phosphorus: 1.37 mmol/L (↑)
    \item Liver/kidney function, uric acid: Within normal range
    \item Rheumatologic markers (anti-CCP, ANA): Within normal range
\end{itemize} \\
\hline
\end{tabular}
\end{table}

\clearpage
\section{Prompt for collecting case analysis from LLMs}\label{AppendixB}
\begin{table}[h]
\centering
\caption{Prompt for collecting case analysis from LLMs in Study One.}
\renewcommand{\arraystretch}{1.5}
\begin{tabular}{p{0.9\textwidth}}
\hline
\textbf{Prompt} \\
\hline

You will play the role of an experienced clinician, conducting a diagnostic consultation. The patient is played by me, based on the medical record of a real but discharged patient. \\

1. \textbf{Consultation Flow:} \\
    - At the beginning, inquire about the patient's history, hypothesize possible disease directions, and then order tests. \\
    - At each turn, you may either ask a question or order an examination. \\

2. \textbf{Asking Questions:} \\
    - If you choose to ask, you cannot order tests in that turn. \\
    - You may ask only one question per turn. \\
    - Keep expressions concise, without markdown formatting or extra explanations. \\

3. \textbf{Ordering Examinations:} \\
    - If you choose to order examinations, you may order at most two tests per turn. \\
    - In that turn, you cannot ask additional questions. \\
    - You may order tests multiple times throughout the consultation. \\
    - Do not repeatedly order examinations that I have already stated were not performed. \\

4. \textbf{Diagnostic Process:} \\
    - Analyze the information to make a diagnosis. \\
    - If multiple possibilities remain, continue asking until incorrect options are excluded. \\
    - If you can make a diagnosis and develop a treatment plan, please conclude the inquiry and present both to me.\\

5. \textbf{Diagnostic Requirements:} \\
    - Provide a clear and precise conclusion including: primary diagnosis, secondary diagnosis, complications, and the rationale for each (based on symptoms, signs, test results, and clinical experience). \\
    - Use standardized medical terminology to ensure accuracy and completeness. \\

6. \textbf{Treatment Plan:} \\
    - Formulate a comprehensive and standardized plan, including pharmacological treatment (drug names, dosages, administration, course of treatment), non-pharmacological interventions, nursing measures, and follow-up plan. \\
    - Following the evidence-based medicine principles, consider patient-specific differences, and ensure safety and effectiveness. \\
\hline
\end{tabular}
\label{tab:prompt}
\end{table}

\clearpage
\section{The correlation between the evaluation dimensions}
\begin{figure*}[h]
    \centering
    \includegraphics[width=\textwidth]{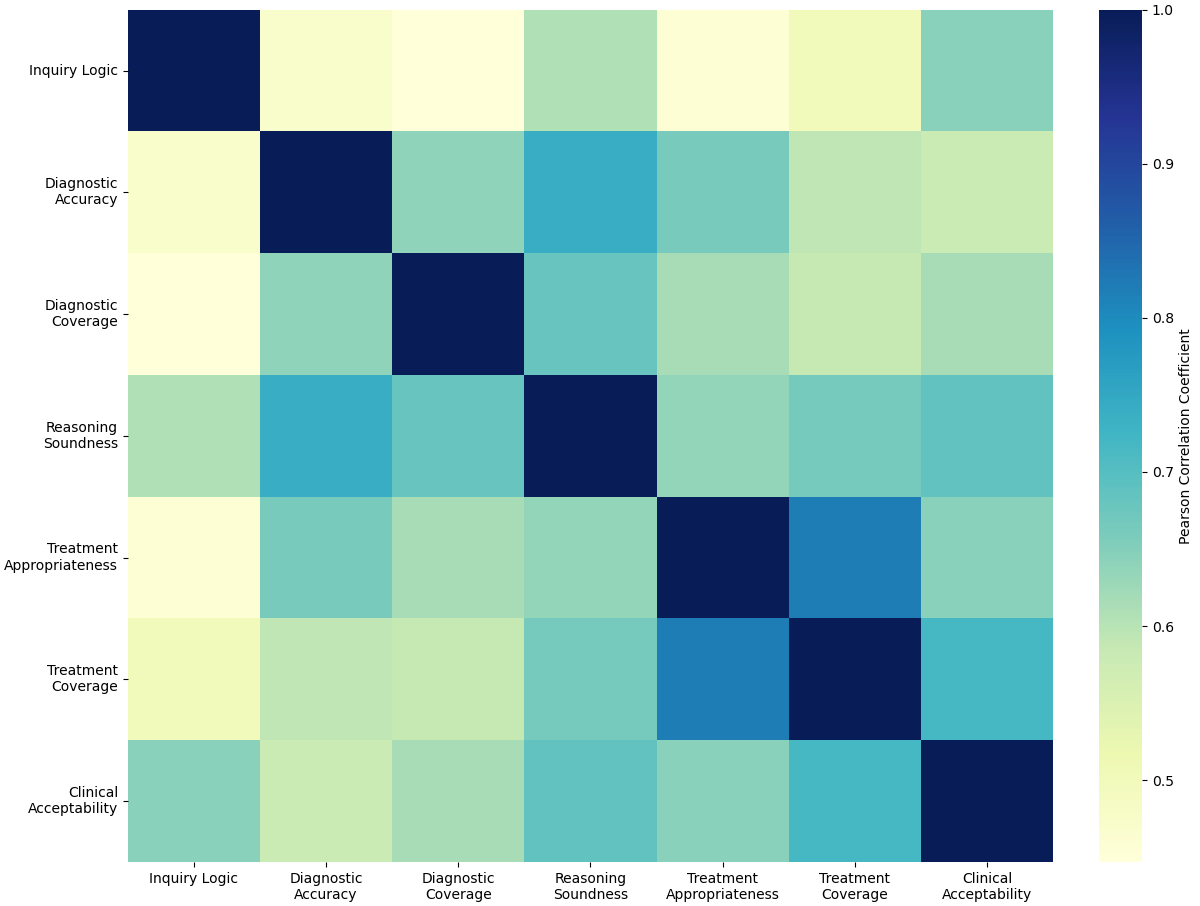}
    \caption{The results of Pearson's correlation coefficient between the evaluation dimensions.}
    \label{fig:dimension_pearson}
\end{figure*}

\clearpage
\section{Estimated threshold parameters from the cumulative link mixed model (CLMM)}
\begin{table}[h]
\centering
\caption{Estimated cutpoints (threshold parameters) from the cumulative link mixed model (CLMM). 
These cutpoints define the thresholds on the latent continuous scale that separate adjacent ordinal categories 
(i.e., category $k$ is chosen when the latent score falls between cutpoint $k-1|k$ and $k|k+1$). 
Smaller (more negative) cutpoint values indicate lower positions on the latent scale. 
(*: $p < 0.05$; **: $p < 0.01$; ***: $p < 0.001$).}
\label{tab:clmm_cutpoints}
\begin{tabular}{lcccc}
\toprule
\textbf{Cutpoint} & \textbf{Estimate} & \textbf{Std. Error} & \textbf{z value} & \textbf{p-value} \\
\midrule
1|2   & -11.77 & 0.50 & -23.48 & $<0.001***$ \\
2|3   & -10.24 & 0.48 & -21.19 & $<0.001***$ \\
3|4   &  -9.05 & 0.47 & -19.28 & $<0.001***$ \\
4|5   &  -8.02 & 0.46 & -17.51 & $<0.001***$ \\
5|6   &  -7.06 & 0.45 & -15.76 & $<0.001***$ \\
6|7   &  -6.10 & 0.44 & -13.87 & $<0.001***$ \\
7|8   &  -5.03 & 0.43 & -11.62 & $<0.001***$ \\
8|9   &  -3.62 & 0.43 &  -8.43 & $<0.001***$ \\
\bottomrule
\end{tabular}
\end{table}

\section{Detailed results for data analysis}\label{AppendixE}
\begin{table}[h]
\centering
\caption{Independent-samples t-test results and Cohen's d for each dimension}
\label{tab:t_test_cohen_d}
\begin{tabular}{cccc}
\hline
\textbf{Dimension} & \textbf{t} & \textbf{p} & \textbf{Cohen's d} \\
\hline
\textit{Inquiry Logic} & -1.44 & 0.19 & -0.52 \\
\textit{Diagnostic Accuracy} & -0.55 & 0.60 & -0.20 \\
\textit{Diagnostic Coverage}  & -0.29 & 0.78 & -0.11 \\
\textit{Reasoning Soundness} & -1.31 & 0.23 & -0.48 \\
\textit{Treatment Appropriateness} & -1.30 & 0.23 & -0.48 \\
\textit{Treatment Coverage}  & -2.08 & 0.07 & -0.76 \\
\textit{Clinical Acceptability} & -1.65 & 0.14 & -0.61 \\
\hline
\end{tabular}
\end{table}

\begin{table}[h]
\centering
\caption{Kendall's W and Friedman $\chi^2$ values for inter-rater consistency of dimension scores within each case (*: p < 0.05; **: p < 0.01; ***: p < 0.001).}
\label{tab:dimension_case}
\begin{tabular}{lcccc}
\toprule
\textbf{Case} & \textbf{Specialty} & \textbf{Friedman $\chi^2$} & \textbf{Kendall's W} & \textbf{p-value} \\
\midrule
1 & Infectious Disease & 51.02 & 0.43 & $<0.001***$ \\
2 & Orthopedics        & 40.34 & 0.34 & $<0.001***$ \\
3 & Orthopedics        & 29.44 & 0.25 & $<0.001***$ \\
4 & Neurology          & 55.52 & 0.46 & $<0.001***$ \\
5 & Neurology          & 41.47 & 0.35 & $<0.001***$ \\
6 & Gastroenterology   & 34.51 & 0.29 & $<0.001***$ \\
7 & Cardiology         & 12.90 & 0.11 & $0.04*$ \\
8 & Gastroenterology   & 36.88 & 0.31 & $<0.001***$ \\
9 & Cardiology         & 52.57 & 0.44 & $<0.05*$ \\
\bottomrule
\end{tabular}
\end{table}

\begin{table}[h]
\centering
\caption{Friedman test and Kendall's W for \textit{Analyst} ranking consistency across cases (*: p < 0.05; **: p < 0.01; ***: p < 0.001).}
\label{tab:Ranking_physician}
\begin{tabular}{p{4cm}p{2cm}p{2cm}p{1.5cm}}
\toprule
\textbf{Analyst} & \textbf{Friedman $\chi^{2}$} & \textbf{Kendall’s W} & \textbf{p-value} \\
\midrule
Deepseek-V3                     & 17.874 & 0.372 & 0.022*        \\
Physician-Specialist (Senior)   & 22.716 & 0.473 & 0.004**       \\
Gemini 2.5 pro                  & 15.593 & 0.325 & 0.049*        \\
GPTo3-mini                      & 15.873 & 0.331 & 0.044*        \\
GPTo3                           & 24.705 & 0.515 & 0.002**       \\
Physician-Specialist (Junior)   & 28.604 & 0.596 & $<$0.001***   \\
Qwen-235B-T                     & 29.138 & 0.607 & $<$0.001***   \\
Qwq-32b                         & 12.608 & 0.263 & 0.126         \\
Non-Specialists                 & 14.665 & 0.306 & 0.66          \\
\bottomrule
\end{tabular}
\end{table}

\begin{table}[h]
\centering
\caption{Friedman and Kendall's W values for \textit{Analyst} ranking differences in each case (*: p < 0.05; **: p < 0.01; ***: p < 0.001).}
\label{tab:case_rank}
\begin{tabular}{lccccc}
\toprule
\textbf{Case} & \textbf{Specialty} & \textbf{Friedman $\chi^2$} & \textbf{Kendall's W} & \textbf{p-value} \\
\midrule
1 & Infectious Disease & 102.17 & 0.64 & $<0.001***$ \\
2 & Orthopedics & 117.10 & 0.66 & $<0.001***$ \\
3 & Orthopedics & 90.69 & 0.44 & $<0.001***$ \\
4 & Neurology & 98.76 & 0.52 & $<0.001***$ \\
5 & Neurology & 74.62 & 0.40 & $<0.001***$ \\
6 & Gastroenterology & 87.82 & 0.51 & $<0.001***$ \\
7 & Cardiology & 111.54 & 0.67 & $<0.001***$ \\
8 & Gastroenterology & 107.29 & 0.67 & $<0.001***$ \\
9 & Cardiology & 62.25 & 0.39 & $<0.001***$ \\
\bottomrule
\end{tabular}
\end{table}

\begin{table}[h]
    \centering
    \caption{\yuansong{Comparison of linear and LOESS regression results.}}
    \label{tab:Regression}
    \begin{tabular}{lcc}
        \hline
        \textbf{Metric} & \textbf{LOESS ($frac=0.7$)} & \textbf{Linear Regression} \\
        \hline
        Coefficient of Determination ($R^2$) & $0.972825$ & $0.896996$ \\
        Residual Standard Deviation (RSD) $\downarrow$ & $3.298667$ & $6.418319$ \\
        Mean Absolute Error (MAE) $\downarrow$ & $2.665099$ & $4.692866$ \\
        \hline
    \end{tabular}
\end{table}

\clearpage
\section{Examples of clinical case analysis (well-performed and poorly-performed)}\label{AppendixF}
\begin{table}[htbp]
\centering
\caption{\yuansong{Diagnostic and treatment process of the \textit{Analysts} ranked highest in \textit{Case 10} (\textit{Gemini2.5 Pro}).}}
\label{tab:Good_example}
\small
\begin{tabular}{p{0.18\textwidth} p{0.78\textwidth}}
\toprule
\textbf{Section} & \textbf{Content} \\
\midrule

\textbf{Conversation} &
\textbf{Standard Patient}: Doctor, I have abdominal pain, abdominal distension, and repeated vomiting. \vspace{2pt} \\
& \textbf{Gemini2.5 Pro}: How long has this been going on? Could you describe the vomitus? Have you passed stool or gas recently? \vspace{2pt} \\
& \textbf{Standard Patient}: 4 days, worsening today; vomitus contains gastric contents and bile; no stool or gas. \vspace{2pt} \\
& \textbf{Gemini2.5 Pro}: Any abdominal surgery history? \vspace{2pt} \\
& \textbf{Standard Patient}: No. \vspace{2pt} \\
& \textbf{Gemini2.5 Pro}: What about abdominal X-ray? \vspace{2pt} \\
& \textbf{Standard Patient}: Not done. \vspace{2pt} \\
& \textbf{Gemini2.5 Pro}: Any tenderness or bowel sounds? \vspace{2pt} \\
& \textbf{Standard Patient}: Epigastric tenderness, reduced bowel sounds. \vspace{2pt} \\
& \textbf{Gemini2.5 Pro}: Enhanced CT and lab results? \vspace{2pt} \\
& \textbf{Standard Patient}: CT: necrotizing pancreatitis, gallstones, pleural effusion. CBC: high WBC and neutrophils. Electrolytes normal. Amylase and lipase elevated. \\

\midrule

\textbf{Diagnosis} &
\begin{minipage}[t]{\linewidth}
1. Severe acute pancreatitis. \\
   Reasoning: Clinical presentation includes acute upper abdominal pain, distension, and vomiting, with reduced bowel sounds, which are typical signs of acute pancreatitis. Biochemical tests show significantly elevated serum amylase and lipase, a key diagnostic criterion. Imaging with enhanced abdominal CT confirms pancreatic inflammation and necrosis, providing direct radiologic evidence. Severity assessment indicates “severe” due to local complications (pancreatic necrosis) and systemic complications (bilateral pleural effusion, suggesting systemic inflammatory response syndrome - SIRS). \\[2pt]
2. Gallstone-induced pancreatitis. \\
   Reasoning: Abdominal CT clearly shows gallstones. In the absence of other obvious causes (e.g., alcohol history), gallstones are the most common etiology of acute pancreatitis. \\[2pt]
3. Pancreatic necrosis. \\
   Reasoning: Identified by CT imaging showing areas of necrotic pancreatic tissue. \\[2pt]
4. Paralytic ileus. \\
   Reasoning: The patient exhibits abdominal distension, cessation of bowel movements, and decreased bowel sounds, indicative of temporary intestinal paralysis caused by severe intra-abdominal inflammation. \\[2pt]
5. Systemic inflammatory response syndrome (SIRS). \\
   Reasoning: Elevated WBC and the presence of pleural effusion indicate that the inflammatory response extends beyond the pancreas, affecting the whole body.
\end{minipage} \\

\midrule

\textbf{Treatment} &
\begin{minipage}[t]{\linewidth}
1. Supportive care: NPO, IV fluids, analgesia, gastric decompression, ICU monitoring. \\
2. ERCP if CBD obstruction is suspected; cholecystectomy after stabilization. \\
3. Early enteral feeding, antibiotics only for infection, organ support, step-up necrosis management.
\end{minipage} \\

\bottomrule
\end{tabular}
\end{table}

\clearpage
\begin{table}[htbp]
\centering
\caption{\yuansong{Diagnostic and treatment process of the \textit{Analysts} ranked lowest in \textit{Case 10} (\textit{Deepseek-V3}).}}
\label{tab:Worst_example}
\small
\begin{tabular}{p{0.18\textwidth} p{0.78\textwidth}}
\toprule
\textbf{Section} & \textbf{Content} \\
\midrule

\textbf{Conversation} & 
\parbox[t]{\linewidth}{
\textbf{Standard Patient}: Doctor, hello, I have abdominal pain, abdominal distension, and repeated vomiting. \\
\textbf{Deepseek-V3}: Could you describe the exact location of your abdominal pain? Also, is the pain dull, colicky, or burning? \\
\textbf{Standard Patient}: The pain is located in the upper abdomen, and it is persistent and dull. \\
\textbf{Deepseek-V3}: What is the nature of your vomitus? Any coffee-ground-like material or blood? Does the pain relate to meals? \\
\textbf{Standard Patient}: Vomitus mainly contains gastric contents; no coffee-ground-like material or blood. Pain is not related to eating. \\
\textbf{Deepseek-V3}: What are your abdominal ultrasound and CBC results? \\
\textbf{Standard Patient}: 1. Abdominal ultrasound: gallstones, cholecystitis, dilated common bile duct. \\
\quad 2. CBC: WBC $16.49 \times 10^{9}/\mathrm{L}$ (high), Neut\% 89.1\% (high), RBC $5.10 \times 10^{12}/\mathrm{L}$, Hb $159\,\mathrm{g/L}$, Platelets $232 \times 10^{9}/\mathrm{L}$.
} \\

\midrule

\textbf{Diagnosis} & 
\parbox[t]{\linewidth}{
1. Conclusion: Acute cholecystitis with gallstones. \\
   Reasoning: 1) Persistent dull pain in the upper abdomen; 2) Vomiting; 3) Ultrasound shows gallstones and cholecystitis; 4) WBC and Neut.\\
2. Conclusion: Dilated common bile duct. \\
   Reasoning: Ultrasound shows dilated common bile duct.
} \\

\midrule

\textbf{Treatment} & 
\parbox[t]{\linewidth}{
1. Pharmacologic treatment: \\
\quad – Ceftriaxone 2g IV once daily (antibiotic) \\
\quad – Metronidazole 0.5g IV every 12 hours (antianaerobic) \\
\quad – Anisodamine 10mg IM PRN (antispasmodic and analgesic) \\
\quad – Omeprazole 40mg IV once daily (acid suppression) \\[2pt]
2. Non-pharmacologic treatment: \\
\quad – NPO (nothing by mouth) \\
\quad – Gastric decompression \\
\quad – IV fluids to maintain water and electrolyte balance
} \\

\bottomrule
\end{tabular}
\end{table}

\end{document}